\documentclass[12pt]{article}

\usepackage{amsmath,amssymb,epsf,setspace}
\usepackage[hang]{caption}
\def\CN{{\cal N}}
\def\CA{{\cal A}}
\def\CO{{\cal O}}
\def\Im{{\rm Im}\,}
\def\tr{{\rm tr}}

\def\footrep{{$^{\thefootnote}$}}

\def\blfootnote{\xdef\@thefnmark{}\@footnotetext} 

\long\def\symbolfootnote[#1]#2{\begingroup%
\def\thefootnote{\fnsymbol{footnote}}\footnote[#1]{#2}\endgroup} 

\newcommand{\nsection}[1]{\setcounter{equation}{0}\section{#1}
  \vspace{5mm}}

\begin{document}

{\flushright April 2012 \\}
\vspace{1cm}

\begin{center}
  \begin{doublespace}
    {\Large \bf \noindent Eikonal Approach to $\CN \!=\! 4$ SYM
      Regge \\ Amplitudes in the AdS/CFT Correspondence}
  \end{doublespace}
  \vspace*{0.5cm}
  {\large Matteo Giordano$^{1,}$\symbolfootnote[1]{\tt giordano@unizar.es}, 
 Robi Peschanski$^{2,}$\symbolfootnote[2]{\tt robi.peschanski@cea.fr} and 
 Shigenori Seki$^{3,2,4,}$\symbolfootnote[3]{\tt sigenori@ihes.fr,
 sigenori@apctp.org}
      }\\
  \vspace*{0.5cm}{\normalsize \it
$^1$ Departamento de F{\' \i}sica Te{\' o}rica,
Universidad de Zaragoza \\
Calle Pedro Cerbuna 12, E-50009 Zaragoza, Spain \\
$^2$ Institut de Physique Th\'{e}orique, CEA-Saclay,
F-91191 Gif-sur-Yvette Cedex, France\\
$^3$ Institut des Hautes {\'E}tudes Scientifiques\\
Le Bois-Marie 35, Route de Chartres, F-91440
Bures-sur-Yvette, France\\
$^4$ Asia Pacific Center for Theoretical
Physics \\ 
San 31, Hyoja-dong, Nam-gu, Pohang 790-784, Republic of Korea
  }
 \vspace*{0.5cm}
\end{center}

\begin{abstract}
  \noindent
The high-energy behavior of $\CN \!=\! 4$ SYM elastic amplitudes 
at strong coupling is studied by means of the AdS/CFT correspondence. 
We consider the eikonal method proposed by Janik and one of the
authors, where the relevant minimal surface
is a ``generalized helicoid'' in hyperbolic space (``Euclidean
$AdS_5$''), from which the physical amplitude is obtained after an
appropriate analytic continuation. 
We then compare our results with those obtained, using a minimal
surface in $AdS_5$ momentum space, by Alday and Maldacena
for gluon-gluon scattering, and 
by Barnes and Vaman for quark-quark scattering  (``Alday-Maldacena
 approach'').  
Exploiting a conformal transformation, we show that the
eikonal amplitudes are dominated by the Euclidean version of the  cusp
contribution found in the Alday-Maldacena approach. 
The amplitudes in the
two approaches are of Regge type at high-energy and with the same
logarithmic Regge trajectory independently of the kind of colliding
particles, in agreement with the expected universality of Regge
trajectories. 

\end{abstract}

\newpage

\nsection{Introduction}

The AdS/CFT correspondence~\cite{Ma} is a powerful non-perturbative tool,
which has been exploited in the study of a variety of problems 
in $\CN \!=\! 4$ supersymmetric Yang-Mills theory (SYM) at strong coupling. In
recent years, a lot of work has been done regarding scattering
amplitudes. In Refs.~\cite{AM,AMi}, Alday and Maldacena have shown
how to obtain the $n$-gluon scattering amplitude in $\CN \!=\! 4$ SYM in
this framework, by finding a minimal surface, corresponding to a
classical string solution, with polygonal boundary in $AdS_5 \times
S^5$. In particular, they have solved analytically the minimal surface
problem in the four-gluon case~\cite{AM}, so obtaining a 
fully analytic expression for the gluon-gluon elastic scattering
amplitude. Their method has been extended to quark-quark scattering in
Refs.~\cite{McS,KR,BV}. In these works, quarks are introduced 
as $\CN \!=\! 2$ hypermultiplets on the field theory side, thus
modifying the dynamical content of the theory, which requires the 
introduction of extra structure, namely D7-branes, in the dual
gravitational description. The D7-branes are then treated in the
probe approximation, neglecting their backreaction, which corresponds
to compute the field theory amplitudes in the quenched approximation,
{\it i.e.}, treating the quarks as external probes. In particular, the
authors of Ref.~\cite{BV} obtain an exact solution to the minimal surface
problem relevant to quark-quark elastic scattering, and although the
area of the surface cannot be expressed in closed form, an explicit
expression can be obtained in the limit of small quark masses.

A different method to compute scattering amplitudes through the
AdS/CFT correspondence had been previously proposed in
Refs.~\cite{JP,JPi,JPii}, in order to evaluate the high-energy
scattering amplitude for 
external quarks. This method is based on the eikonal approximation and
the Wilson-line formalism for high-energy
amplitudes~\cite{Na,DFK,Me,Mei}, and on analytic continuation to
Euclidean space~\cite{Meii,Meiii,Meiv,GM}. 
In this case, no new dynamical degree of freedom is added on the
field theory side, so that no extra structure has to be introduced in
the dual gravitational description. Quarks are treated directly from
the onset as external particles coupled to the gauge (and scalar) fields
of the $\CN=4$ theory. In this approach, the scattering amplitude is
obtained from the correlation function of two Wilson lines running
along the eikonal trajectories of the quarks. Through analytic
continuation and gauge/gravity duality, this correlation function is
related to the area of a minimal surface in Euclidean $AdS_5$ ({\it
i.e.}, hyperbolic space), whose boundaries are two straight lines,
corresponding to the Euclidean trajectories of the quarks. 
The relevant minimal surface thus corresponds to a  ``generalized
helicoid''~\cite{JPi} in the AdS background, characterized by the
impact-parameter distance between the quarks and by the opening angle
$\theta$ of the boundary. After analytic continuation back into
Minkowski space, one obtains the impact-parameter amplitude at given
high enough rapidity $\chi$. However, the expressions obtained in
\cite{JPi} were not complete, suffering from the lack of knowledge on the
exact analytic form of the ``generalized helicoid''. One goal of the
present paper is to go further in the eikonal approach, in order to go
beyond the approximations made in~\cite{JPi}, and so obtain a more refined
result.

One major interest of the eikonal method is that it can be extended to
non-conformal backgrounds~\cite{JPi,JPii,GPi}, corresponding to generic
non-conformal gauge field theories, where using general features of
gauge/gravity duality it leads in this case to Regge amplitudes with
linear trajectory. Our aim in the present study is to look for
Regge behavior of amplitudes in the conformal case of $\CN \!=\! 4$ SYM by
using this method.

Indeed, the high-energy behavior of scattering amplitudes has been
analyzed for a long time in terms of Regge amplitudes, both from the 
phenomenological and the theoretical point of view (see {\it
  e.g.}~Ref.~\cite{Col}).  
As it is well known, the Regge behavior is a remarkable property of
Yang-Mills theories in the perturbative regime. However, the issue of
Regge behavior of high-energy amplitudes at strong coupling requires
different tools, and the AdS/CFT correspondence seems to be well
suited for this purpose. In the case of gluon-gluon scattering, the
analysis of the high-energy 
behavior has been carried out in Refs.~\cite{NS,DKS}, based on the
Alday-Maldacena result of Ref.~\cite{AM}, and on dual conformal
symmetry~\cite{DHSS} and the all-order BDS ans\"atz
of Ref.~\cite{BDS}, showing indeed the Regge nature of the amplitude
(which in particular is Regge-exact in the $s$-channel~\cite{DKS}). 
This analysis can be easily extended to the results of Ref.~\cite{BV}, which
will allow us to discuss the issue of universality of Regge amplitudes
in $\CN \!=\! 4$ SYM  at strong coupling.  On the
other hand, the comparison with the results obtained in the eikonal
approach allows to  check the compatibility of the two methods, which
are based on very different constructions, and thus provide a
nontrivial test for the validity of the eikonal approach. This is very
important in view of the application of the eikonal approach to QCD,
where an analogue of the Alday-Maldacena approach is not currently available,
and moreover allows to look at the universality problem in a different
way.

The plan of the paper is the following.
In Section 2, we give a brief review of the two methods 
for approaching the high-energy behavior of $\CN \!=\! 4$ SYM amplitudes,
namely the Alday-Maldacena approach of Ref.~\cite{AM}, and the eikonal approach
of Ref.~\cite{JPi}. In Section 3, we investigate in detail the minimal
surface related to gluon-gluon scattering obtained by Alday and
Maldacena. In particular, the IR boundary of this solution is
analyzed, together with the UV boundary of a corresponding solution in
Euclidean $AdS_5$, generated by analytic continuation.   
In Section 4, we investigate the high-energy domain of the
Alday-Maldacena gluon-gluon
scattering amplitude, both in the momentum and in the impact-parameter
representation, making explicit that in this domain the amplitude is
of Regge type. Moreover, we compare the result with the quark-quark
scattering amplitude of Ref.~\cite{BV}, and discuss the issue of
universality of the Regge trajectory.
In Section 5, we study the minimal surface problem in Euclidean
$AdS_5$ relevant to quark-quark scattering in the eikonal method in a
new way, which allows us to go beyond the preliminary 
results of Ref.~\cite{JPi}. In particular, we show that the amplitude is of
Regge type, and we obtain the leading behavior of
the Regge trajectory, which we show to be in agreement with the
trajectory obtained with the Alday-Maldacena method. We also extend the results
of the eikonal approach to the gluon-gluon scattering case, finding
the agreement expected in the light of universality. Finally, Section
6 is devoted to conclusions and outlook.

\nsection{Two-body elastic scattering {\it via} the AdS/CFT correspondence}

\subsection{The Alday-Maldacena approach}

The gluon four-point scattering amplitude in $\CN \!=\! 4$ SYM has been
evaluated in Ref.~\cite{AM}, making use of the AdS/CFT correspondence, by
computing the area of a corresponding minimal surface. 
In the dual gravity theory, which is defined in $AdS_5\times S^5$, 
the gluon-gluon scattering amplitude is mapped into the scattering amplitude 
of four open strings. 
In turn, the string amplitude is obtained by determining 
a minimal surface, corresponding to a classical string solution 
for the Nambu-Goto action. 
This minimal surface lives in the $AdS_5$ background,
\begin{equation}
  \label{metorig}
  ds^2 = {R^2 \over z^2}\big(\eta_{\mu\nu}dx^\mu dx^\nu + dz^2\big) \,,
\end{equation}
where $\mu = 0,1,2,3$ and $\eta_{\mu\nu} = {\rm diag}(-1,1,1,1)$.
We call this background 
the {\it position space}. 
The idea of Ref.~\cite{AM} is to find the minimal surface 
in {\it momentum space}, rather than directly in the position space. 
The momentum space $(y^\mu,r)$ is obtained from the position space 
$(x^\mu,z)$ by means of the T-duality transformation,
\begin{equation}
  \label{Tdual}
\partial_m y^\mu = i {R^2 \over z^2} \epsilon_{mn}\partial_n x^\mu \,,  
\end{equation}
and the resulting metric is given by
\begin{equation}
  \label{metTdual}
ds^2 = {R^2 \over r^2}\big(\eta_{\mu\nu}dy^\mu dy^\nu + dr^2 \big) \,, \quad 
r \equiv {R^2 \over z} \,.  
\end{equation}
In the momentum space, the boundary of the minimal surface corresponding
to the four-gluon amplitude ({\it i.e.}~to two-body scattering) is 
given by the closed sequence of four light-like segments $\Delta y_i^\mu$. 
The boundary conditions in the position space, {\it i.e.}, 
that the vertex-operator insertion point $x_i$ carries the momentum $k_i$ 
of the corresponding open string, 
translates into the condition $\Delta y_i=2\pi k_i$. 
In the same way, the gluon $n$-point amplitude is
obtained from the minimal surface having as boundary a closed sequence
of $n$ light-like segments \cite{AMi}. 
The sequences are closed because of momentum conservation. 
The light-like segments lie at $r=r_{\rm IR}=R^2/z_{\rm IR}$,
where $z_{\rm IR}$ is the fifth coordinate in position space of the D-brane 
on which the open strings end. 
Such a D-brane acts as a regulator for the IR divergencies 
of the gluon-gluon scattering amplitude, which has to be
removed by sending $z_{\rm IR}\to\infty$, {\it i.e.}, $r_{\rm IR}\to 0$, 
at the end of the calculation. 
It is however more convenient to find the minimal surface directly 
at $r_{\rm IR}=0$, which requires to trade $r_{\rm IR}$ 
for a different IR regulator when evaluating the area of the surface.

The solution obtained in Ref.~\cite{AM} for the minimal surface relevant to
the  gluon four-point scattering amplitude reads in momentum space 
\begin{subequations}
\label{solTdual}
\begin{align}
y_0 &= {\alpha\sqrt{1+\beta^2} \sinh u_1 \sinh u_2 \over \cosh u_1
\cosh u_2 + \beta \sinh u_1 \sinh u_2} \,,\label{solTdual{a}} \\ 
y_1 &= {\alpha \sinh u_1 \cosh u_2 \over \cosh u_1 \cosh u_2 + \beta
\sinh u_1 \sinh u_2} \,,  \label{solTdual{b}} \\ 
y_2 &= {\alpha \cosh u_1 \sinh u_2 \over \cosh u_1 \cosh u_2 + \beta
\sinh u_1 \sinh u_2} \,,  \label{solTdual{c}} \\ 
y_3 &= 0 \,,  \label{solTdual{d}} \\ 
r &= {\alpha \over \cosh u_1 \cosh u_2 + \beta \sinh u_1 \sinh u_2}
\,, \label{solTdual{e}} 
\end{align}
\end{subequations}
where $u_{1,2}$ are world-sheet coordinates 
on the surface ranging from $-\infty$ to
$+\infty$. 
The parameters $\alpha, \beta$ are related to the Mandelstam
variables\footnote{The Mandelstam variables are defined here by 
  \begin{align*}
&-s = (k_1 + k_2)^2 = 2k_{1\mu} k_2{}^\mu \,, \quad
-t = (k_1 + k_4)^2 = 2k_{1\mu} k_4{}^\mu \,, \\ 
&-u = (k_1 + k_3)^2 = 2k_{1\mu} k_3{}^\mu = s+t \,.
  \end{align*}
Note that the physical scattering region that we are considering here 
is $s,t <0$ and $u>0$, which is called the ``$u$-channel'' in the 
literature. Moreover, in the Regge region one has $u \gg 1$ and $t$
fixed, so that $-s \sim u$.  
} 
$s,t$ as 
\begin{equation}
  \label{mandpara}
-s\,(2\pi)^2 = {8\alpha^2 \over (1-\beta)^2} \,, \quad 
-t\,(2\pi)^2 = {8\alpha^2 \over (1+\beta)^2} \,.
\end{equation}
By the use of the T-dual transformation \eqref{Tdual}, 
the minimal surface \eqref{solTdual} is mapped back into the
position space as   
\begin{subequations}
  \label{solorig}
  \begin{align}
    x_0 &= {iR^2 \over 2\alpha}\sqrt{1+\beta^2}(\cosh^2 u_2 - \cosh^2 u_1)
    \,, \label{solorig{a}} \\
    x_1 &= {iR^2 \over \alpha}\bigg[ {u_2 \over 2} +{1 \over 4}\sinh 2u_2
    +\beta\bigg(-{u_1 \over 2} +{1 \over 4}\sinh 2u_1 \bigg)\bigg] \,,
    \label{solorig{b}} \\
    x_2 &= {iR^2 \over \alpha}\bigg[ -{u_1 \over 2} -{1 \over 4}\sinh
    2u_1 +\beta\bigg({u_2 \over 2} -{1 \over 4}\sinh 2u_2 \bigg)\bigg]
    \,,\label{solorig{c}} \\ 
    x_3 &= 0 \,, \label{solorig{d}}\\ 
    z &= {R^2 \over \alpha}(\cosh u_1 \cosh u_2 +\beta \sinh u_1 \sinh
    u_2) \,. \label{solorig{e}} 
  \end{align}
\end{subequations}
Substituting the minimal surface solution \eqref{solTdual} into the
Nambu-Goto action, the gluon-gluon scattering amplitude is evaluated as   
\begin{align}
\CA^{\rm gluon} &= e^{iS} = \exp \bigg[2iS_{\rm div}(s) +2iS_{\rm div}(t)
+{\sqrt{\lambda} \over 8\pi}\bigg(\log{s \over t}\bigg)^2 +{\tilde
C}\bigg] \,, \label{ampAM} \\ 
&iS_{\rm div}(p) = -{1 \over \epsilon^2}{1 \over 2\pi}\sqrt{\lambda
\mu^{2\epsilon} \over (-p)^\epsilon} -{1 \over \epsilon}{1 \over
4\pi}(1-\log 2)\sqrt{\lambda \mu^{2\epsilon} \over (-p)^\epsilon} \,,
\quad (p=s,t) \label{ampAMdiv} 
\end{align}
where ${\tilde C}$ is a constant that is irrelevant to our purposes. 
Here $\lambda$ is the 't Hooft coupling defined
by $\sqrt{\lambda} \equiv \sqrt{g_{\rm YM}^2 N_c} = R^2/\alpha'$,
and we have adopted units where $\alpha'=1$. 
Dimensional regularization has been employed 
in order to obtain a finite result for the area of the minimal surface, 
by going to $D=4-2\epsilon$ dimensions (with $\epsilon<0$). 
This requires the introduction of an IR cutoff scale $\mu$, 
having dimensions of mass, to account for the mass dimension of
the $D$-dimensional coupling. 
Note that the expression \eqref{ampAM} agrees with the BDS ans{\" a}tz
\cite{BDS}  in the
strong coupling limit.

The approach of Ref.~\cite{AM} has been extended to the case of
quark-quark scattering in Refs.~\cite{McS,KR,BV}. The scattering
amplitude is related to a minimal surface in a modified gravitational 
background including D7-branes, whose positions in the radial
direction of AdS corresponds to the masses of the various flavours of
quarks. In particular, Ref.~\cite{BV} provides an exact solution, although
in implicit form, for the minimal surface relevant to elastic quark-quark
scattering. An explicit expression for the regularized area is also
obtained in the limit of small quark masses, which we will report in
Section 4.

\subsection{The eikonal approach}

Let us recall now some relevant elements of the derivation of the
quark-quark elastic scattering amplitude in the high-energy domain, 
in the framework of the AdS/CFT correspondence, 
following the eikonal approach of Ref.~\cite{JPi}. 
The starting point is the formulation of high-energy elastic scattering
amplitudes, 
at fixed and small momentum transfer,\footnote{In the original 
formulation~\cite{Na}, valid for QCD, ``small'' means that the momentum
transfer $t$ has to be smaller than the typical hadronic scale,
$|t|\lesssim 1 \,{\rm GeV}^2$. Since we are dealing here 
with a conformal theory, ``small'' can only mean that it has to be
smaller than the center-of-mass total energy squared $s$, {\it i.e.},
$t\ll s$. Moreover, Wilson lines include the contribution of scalar
fields to the non-Abelian phase factor, as explained in Ref.~\cite{Mai}.
}
in terms  of the correlation function $\CA^{qq}$ 
of two Wilson lines~\cite{Na,DFK,Me,Mei},
\begin{equation}
  \label{corrfuncdef}
\CA^{qq}_{ij,kl} = {1\over Z_W^2}\langle
(W_1-1)_{ij}(W_2-1)_{kl}\rangle\,, \quad Z_W = {1\over N_c}\langle
 {\rm tr}\, W_1 \rangle = {1\over N_c}\langle
 {\rm tr}\, W_2 \rangle \,,
\end{equation}
where $Z_W$ is a renormalisation constant, which makes $\CA^{qq}$
UV-finite (see $e.g.$~\cite{Meii}). 
The relevant Wilson lines $W_{1,2}$ run along infinite light-like
straight lines, at transverse separation $b$, and are taken in the
representation appropriate for the particles under consideration. 
We will be interested initially in the scattering of massive quarks
(antiquarks) in the fundamental (anti-fundamental) representation,  
which we use as external probes of $\CN \!=\! 4$ SYM. 
This approach essentially amounts to consider the eikonal approximation 
for the elastic amplitude, which is expected to be valid
in the Regge kinematic region for $\CN \!=\! 4$ SYM (as well as for
QCD).

The correlation function \eqref{corrfuncdef} yields the impact-parameter 
representation for the elastic scattering amplitudes in the $s$-channel. 
In order to regularize IR divergencies, the Wilson lines are cut at
some proper time $\pm T$, and moved slightly away from the light-cone. 
In this way, they correspond to the classical trajectories 
of two massive quarks, which form a finite hyperbolic angle $\chi$, 
related to the center-of-mass total energy squared $s$ 
as $\chi \sim \log (s/M^2)$ at high energy. 
Here $T$ acts as an IR regulator, which has to be removed at the end
of the calculation by taking the limit $T\to\infty$, 
while the quark mass $M$ is irrelevant in the large $\chi$ region. 
Explicitly, the quark-quark
scattering amplitude is then given by~\cite{Na,DFK,Me,Mei}
\begin{equation}
  \label{eikampldef}
{\cal M}^{qq\,\,\alpha'\!\alpha,\beta'\!\beta}_{ij,kl}(s,t) =
\delta_{\alpha'\!\alpha}\delta_{\beta'\!\beta}{\cal
M}^{qq}_{ij,kl}(s,t) =  
\delta_{\alpha'\!\alpha}\delta_{\beta'\!\beta} (-2is)\int d^2 
b\,e^{i\vec q\cdot\vec b} 
\CA^{qq}_{ij,kl}(\chi,b,T)\,,
\end{equation}
where $\alpha',\alpha$ (resp.~$\beta',\beta$) are the final and initial
spin indices of quark 1 (resp.~2), and $i,j$ (resp.~$k,l$) are the
final and initial color indices of quark 1 (resp.~2). 
Moreover, $\vec q$ and $\vec b$ are two-dimensional vectors in the
transverse plane, with $t=-\vec q^{\,2}$ and $b=|\vec b|$. Here the
limits $\chi\to\infty$, $T\to\infty$ are understood.

It has been shown that the Minkowskian Wilson line correlation function 
$\CA^{qq}(\chi,b,T)$ can be reconstructed 
from the correlation function $\CA^{qq}_E(\theta,b,T)$ 
of two corresponding Euclidean Wilson lines, 
by means of analytic continuation \cite{Meii,Meiii,Meiv,GM}. 
The relevant Euclidean Wilson lines run along straight lines of length $2T$,
which form now an angle $\theta$ in Euclidean space, 
and are separated by the same transverse distance $b$ as in the
Minkowskian case.  
Starting from $\CA^{qq}_E$, the quark-quark elastic scattering amplitude 
in the $s$-channel is obtained by means of 
the analytic continuation relation \cite{Meiv}, 
\begin{equation}
  \label{ancontrel}
\CA^{qq}_{ij,kl}(\chi,b,T) = \CA_{E\, ij,kl}^{qq}(-i\chi,b,iT)\,, \quad 
\CA_{E\, ij,kl}^{qq}(\theta,b,T)=\CA^{qq}_{ij,kl}(i\theta,b,-iT) \,. 
\end{equation}
Moreover, the impact-parameter amplitude in the crossed $u$-channel
$\CA^{q{\bar q}}$, corresponding to quark-antiquark scattering 
at center-of-mass energy squared $u$ ($u>0$), 
can be obtained through the crossing-symmetry relations \cite{GM}
\begin{equation}
  \label{crossrel}
\CA^{q\bar{q}}_{ij,kl}(\chi,b,T) = \CA^{qq}_{ij,lk}(i\pi-\chi,b,T) 
	= \CA_{E\, ij,lk}^{qq}(\pi+i\chi,b,iT) =
        \CA_{E\, ij,kl}^{q\bar{q}}(-i\chi,b,iT) \,,   
\end{equation}
where in the last passage 
$\CA_{E\, ij,kl}^{q\bar{q}}(\theta,b,T) \equiv \CA_{E \,
ij,lk}^{qq}(\pi-\theta,b,T)$  
is the crossed Euclidean amplitude, 
and where $\chi$ has to be identified with 
\begin{equation}
  \label{Quarkchi}
\chi \sim \log{u \over M^2} \sim \log{-s \over M^2}  
\end{equation}
in the high-energy limit.\footnote{It is easy to see that the transformation 
$\chi\to i\pi-\chi$ (with $\chi>0$) corresponds 
to $s\to e^{-i\pi}u$ (with $s,u>0$) in terms of Mandelstam variables.
} 
This relation will be useful further on, when comparing 
with the Alday-Maldacena amplitude. 

The Euclidean Wilson-line correlation functions can be computed 
through the AdS/CFT correspondence, following the approach of
Ref.~\cite{Mai}.  
On the field theory side, the fundamental Wilson lines running 
along straight lines describe the propagation of heavy fundamental
particles in Euclidean space. Using the gauge invariance of the
vacuum, the Euclidean correlation function $\CA^{qq}$ can be
decomposed into a singlet and an ``octet'' part,
\begin{equation}
  \label{colordec}
\CA^{qq}_{E\,ij,kl} = \CA_0\delta_{ij}\delta_{kl} +
\CA_{N_c^2-1}t^a_{ij}t^a_{kl}\,,  
\end{equation}
where $t^a$ are the generators of $SU(N_c)$ in the fundamental
representation, and simple algebra allows to relate the coefficients
of the two color 
structures to the (normalized) expectation values of the Wilson loops
obtained by properly closing the contour at infinity, namely\footnote{In
order to make the equations more transparent, we have preferred to
substitute the exact expression of the subtraction constant $-2Z_W +
1$ with its value $-1$ obtained through the AdS/CFT correspondence,
where $Z_W\sim 1$, see below.}
\begin{align}
\CA_0 &= {1\over Z_W^2}\bigg({1\over N_c^2} \langle {\rm tr}\, W_1 {\rm tr}\,
W_2\rangle -  1\bigg)
\,,\label{colordectwo}\\
\CA_0 + {N_c^2-1 \over 2N_c} \CA_{N_c^2-1} &= {1\over Z_W^2}\bigg({1\over N_c}
\langle {\rm tr}\, W_1  W_2\rangle - 1\bigg) \,. \label{colordecthree}
\end{align}
We stress the fact that there is no relation between the heavy
particles in Euclidean space and the ``physical'' quarks in Minkowski
space: indeed, the Euclidean particles are only introduced as an
intermediate device to compute the relevant Wilson-loop expectation
values, playing no role in the physical process under consideration. We
will return on this point in the following.

Massive particles can be introduced in $\CN \!=\! 4$ SYM 
by breaking the $SU(N_c+1)$ symmetry to $SU(N_c)\times U(1)$, 
which gives rise to massive $W$-bosons
transforming in the fundamental representation of $SU(N_c)$.  
On the gravity theory side, this can be accomplished 
by stretching one of the $N_c+1$ branes away from the others, 
and towards the boundary $z=0$ of Euclidean $AdS_5$. 
The mass of the $W$-bosons is related to the position $z_B$ 
of the displaced brane as $M_B\sim z_B^{-1}$, 
and therefore it becomes very large as $z_B \to 0$. 
The Wilson loop describing the propagation of the $W$-bosons 
along a closed contour ${\cal C}$ is identified in the dual bulk theory 
as the partition function of a string propagating in Euclidean $AdS_5$, 
with the boundary condition that it ends on the contour ${\cal C}$ 
at the boundary $z=0$. 
To leading order, it is therefore given by 
$\langle {\cal W} \rangle \sim \exp(-A)$ 
with $A$ the (properly regularized) area\footnote{Note that the factor
$\sqrt{\lambda}/(2\pi)$ is included into the area $A$.} 
of a minimal surface in Euclidean $AdS_5$, 
ending on ${\cal C}$ at the boundary $z=0$. 
Also in this case it is convenient to work directly in the limit $z_B=0$, 
while at the same time regularizing the area (in the UV) 
by limiting the integration to the region $z>\epsilon$. UV
divergencies are dealt with by means of the Legendre transform
prescription of Ref.~\cite{DGO}.

A remark is in order here. 
Since we are considering heavy (Euclidean) particles, 
the boundary conditions for the minimal surface in the supergravity description 
of the problem are naturally given at the UV, $z=0$. 
This is in contrast with the calculation of Ref.~\cite{AM}, 
where such boundary conditions are given at the IR, $z=\infty$, 
which is again natural for massless particles. 
One question we want to answer to is 
how the two points of view can be reconciled. 
Let us note that while the computation of Ref.~\cite{AM} is
performed in Minkowski space, 
here we are considering a calculation in Euclidean space, 
from which the physical, Minkowskian result for the
scattering amplitude is recovered only after analytic continuation. 
In Euclidean space, the heavy $W$-boson 
is introduced only to establish a connection between the expectation
value of the relevant Wilson loops on the field theory side, and their
dual description on the gravity side. In particular, the $W$-boson mass plays
only the role  of a UV regulator in the computation of the area of the
relevant minimal surfaces, and drops from the 
Wilson-loop expectation values 
after UV divergencies have been removed, {\it before} the analytic
continuation. 
On the other hand, the physical scattering amplitude depends on the
mass $M$ of the Minkowskian quarks (and on $s$) only through the 
dependence of the relevant (Minkowskian) Wilson-loop expectation values
on the hyperbolic angle $\chi$. In other words, the dependence on $M$
appears only when the relation between $\chi$, $s$ and $M$ 
is made explicit {\it after} the analytic continuation. 
This shows that the mass of the Euclidean (heavy) $W$-boson 
and the mass of the Minkowskian quarks are completely unrelated.  
We see therefore that there is a natural connection between the use of
very heavy particles in Euclidean space, and the final goal of describing
the scattering of particles with very high energy in Minkowski space,
the link being provided by the use of Wilson loops and by the analytic
continuation \eqref{ancontrel}. 

We specialize now to the case of interest, {\it i.e.}, 
the Euclidean correlator $\CA^{qq}_E(\theta,b,T)$. First of all, we
notice that the normalisation factor reduces to $Z_W\sim 1$, due to
the Legendre transform prescription~\cite{DGO}. This prescription implies
also that the disconnected contributions to the expectation values on
the right hand side of Eqs.~\eqref{colordectwo}
and~\eqref{colordecthree} is 1, and 
therefore gets cancelled. Next, since the connected part of
$\CA_0$ is related to a minimal surface with tube topology, we have that
$\CA_0 = {\cal O}(1/N_c^2)$, and so the amplitude is dominated by
the ``octet'' component $\CA_{N_c^2-1}$, which is of order ${\cal
O}(1/N_c)$: indeed, the relevant minimal surface in
Eq.~\eqref{colordecthree} has disk topology, so that the right hand side is
of order 1. Finally, we obtain at large $N_c$
\begin{equation}
  \label{amplisurf}
\CA^{qq}_{E\,ij,kl} \sim \CA_{N_c^2-1}t^a_{ij}t^a_{kl} \sim {1\over
  N_c} \delta_{il}\delta_{jk}{1\over N_c}
\langle {\rm tr}\, W_1  W_2\rangle_c \equiv {1\over
  N_c} \delta_{il}\delta_{jk} \tilde{\CA}^{\rm quark}_E\,,
\end{equation}
where the subscript $c$ stands for the connected component.

The basic building block of the construction is therefore a minimal
surface in anti-de Sitter space, which is bounded by
two oriented straight 
lines at the boundary $z=0$ of $AdS_5$, corresponding to the
trajectories of the two heavy Euclidean quarks in the static (infinite
mass) limit. We call this surface a ``generalized helicoid''. 
\begin{figure}[t]
  \centering
  \epsfbox{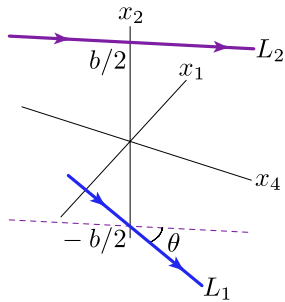}
  \caption{The two straight-line trajectories defining
  the UV boundary of the minimal surface in the eikonal
  approach.}
  \label{figquarkpara}
\end{figure}
In order to properly define the variational problem, 
it is convenient to take the two lines to have infinite length, 
while at the same time introducing a new IR cutoff to regularize 
the area of the resulting minimal surface. 
In practice, the boundary is defined by the two straight
lines\footnote{We use the convention $(x_1,x_2,x_3,x_4)$ for the
  coordinates of points in Euclidean space.
} 
\begin{equation}
  \label{lines}
L_1:\bigg(-\tau \sin{\theta \over 2},-{b \over 2},0,\tau \cos{\theta
\over 2}\bigg) \,, \quad 
L_2:\bigg(\tau \sin {\theta \over 2},{b \over 2},0,\tau \cos{\theta
\over 2}\bigg) \,, \qquad  
-\infty \leq \tau \leq \infty \,,  
\end{equation}
traveled from $\tau=-\infty$ to $\tau=+\infty$, 
separated by a distance $b$ in the ``transverse'' direction $x_2$, and
forming a relative angle $\theta$ in the ``longitudinal'' plane
$(x_1,x_4)$ (see \ref{figquarkpara}). On the other hand, since the
area functional
\begin{equation}
  \label{areafunc}
A^{\rm quark}_{\theta,b} = {\sqrt{\lambda} \over 2\pi} 
	\int d\tau d\sigma {1 \over z^2}
	\sqrt{\det\big({\delta_{\mu\nu}\partial_a x^\mu \partial_b x^\nu
	+ \partial_a z \partial_b z}\big)}  
	\equiv \int d\tau d\sigma {\cal L}  
\end{equation}
with the boundary \eqref{lines} at $z=0$ is expected to be
infinite due to IR divergences,\footnote{The quantity $A^{\rm quark}$ 
is also UV divergent due to the behavior of the metric near the boundary $z=0$, 
the divergence taking the form 
$A^{\rm quark}_{\rm UV\,\,div}= \sqrt{\lambda}(2\pi)^{-1}{4T / \epsilon}$.
This is precisely the area of two planar ``walls'', extending along
the Wilson lines and in the fifth dimension of AdS, which is
subtracted from the minimal area when using the Legendre prescription
of Ref.~\cite{DGO}.} 
we limit the range of $\tau$ to $\tau\in[-T,T]$, 
understanding that it has to be imposed in the computation of the area, 
and not in the determination of the minimal surface. 
The regularized (and UV-subtracted) area of the surface minimizing 
the functional \eqref{areafunc} is therefore a function 
$A^{\rm quark}_{\rm min}(\theta,b,T)$, 
which enters the color-independent part of the Euclidean
``amplitude'', defined in Eq.~\eqref{amplisurf}, as 
$\tilde{\CA}_E^{\rm quark}(\theta,b,T) 
= \exp\big[-A^{\rm quark}_{\rm min}(\theta,b,T)\big]$. 
This function will be defined more precisely in Section 5.1.

The case of quark-antiquark scattering is obtained 
by simply flipping the orientation of one of the two straight lines,
{\it e.g.},
\begin{equation}
  \label{antiq}
L_2 \to L'_2:\bigg(-\tau \sin {\theta \over 2},{b \over 2},0,-\tau
\cos{\theta \over 2}\bigg) \,,   
\end{equation}
with $\tau$ running again from $-\infty$ to $+\infty$. 
This corresponds to changing the representation 
of the corresponding Wilson line from fundamental to anti-fundamental, 
as appropriate for an antiquark. 
In turn, exploiting the Euclidean symmetries, 
it is easy to see that this is equivalent 
to the change $\theta\to \pi -\theta$ in the relative angle. 
The minimal surface relevant to quark-antiquark scattering 
is therefore obtained by minimizing $A^{\rm quark}_{\pi-\theta,b}$, 
and thus it is equal to $A^{\rm quark}_{\rm min}(\pi-\theta,b,T)$, 
so that the two cases can be treated at once.

In the case of the $AdS_5\times S^5$ background 
one does not know yet the minimal surface 
corresponding to the boundaries \eqref{lines}. 
A simple scheme has been introduced in Ref.~\cite{JPi}, 
where the following ans{\" a}tz for the ``generalized helicoid''
is assumed in order to find the minimal solution,\footnote{This 
ans{\" a}tz~\cite{JPi} corresponds to a conjectured generalization of the usual
Euclidean helicoid to the AdS metric. Although the exact solution is
not necessarily parameterizable in the same way, we nevertheless expect 
this ans{\" a}tz to be reasonable, and at least a controllable
approximation of the exact solution.}  
\begin{equation}
  \label{helico}
 x_1 =\tau \sin {\theta \sigma \over b} \,, \quad
x_2 =\sigma \,, \quad 
x_3 = 0 \,, \quad
x_4 =\tau \cos {\theta \sigma\over b} \,, \quad
z =z(\tau, \sigma) \,. 
\end{equation}
The world-sheet coordinates $\tau,\sigma$ are 
in the range, $\tau \in [-\infty,\infty]$ and $\sigma \in [-b/2,b/2]$. 
Using this ans{\" a}tz, the regularized area functional
\eqref{areafunc} becomes   
\begin{equation}
  \label{JParea}
 A^{\rm quark}_{\pi-\theta,b} = {\sqrt{\lambda} \over 2\pi} \int_{-T}^{T} d\tau \int_{-b/2}^{b/2}
d\sigma {1 \over z^2} \sqrt 
{ \left(1+{{\tau^2\theta^2}\over{b^2}}\right) \big(1+(\partial_\tau
z)^2\big) +(\partial_\sigma z)^2 } \,,
\end{equation}
where the IR cutoff parameter $T$ is introduced, as explained above.

We remark here that the ans{\" a}tz \eqref{helico} is appropriate 
for quark-antiquark scattering, that is, for the correlator 
$\CA^{q\bar q}_E(\theta,b,T)=\CA^{qq}_E(\pi-\theta,b,T)$, 
rather than for quark-quark scattering. 
The reason is that if we want an orientable surface, 
the two straight-lines which form the boundary of the helicoid 
have to be travelled in opposite directions, 
if the surface performs a twist of angle $\theta$. 
On the other hand, if they are travelled in the same direction, 
the helicoid has to perform a twist of angle $\pi-\theta$ 
in order to obtain an orientable surface. 
For this reason, we have denoted as $A^{\rm quark}_{\pi-\theta,b}$ 
the area functional in Eq.~\eqref{JParea}. 
Nevertheless, as explained above, the geometrical problem 
to be solved in Euclidean space is the same for quark-quark and
quark-antiquark scattering. 
The difference between the two cases lies in the specific analytic continuation 
which one has to make in order to obtain the physical amplitude. 

We conclude this section with a brief description of the treatment of
gluon-gluon scattering in the eikonal method. The gluon-gluon
scattering amplitude is given by the expression
\begin{equation}
  \label{eikampldefgg}
{\cal M}^{gg}_{ab,cd}(s,t) =  
-2is\int d^2 
b\,e^{i\vec q\cdot\vec b} 
\CA^{gg}_{ab,cd}(\chi,b,T)\,,  
\end{equation}
up to helicity-conserving Kronecker deltas, with
\begin{equation}
  \label{corrfuncdefgg}
\CA^{gg}_{ab,cd} = {1\over Z_V^2}\langle
(V_1-1)_{ab}(V_2-1)_{cd}\rangle\,, \quad Z_V = {1\over N_c^2-1}\langle
 {\rm Tr}\, V_1 \rangle = {1\over N_c^2-1}\langle
 {\rm Tr}\, V_2 \rangle \,.  
\end{equation} 
Here $V_i$ are Wilson lines in the adjoint representation, running on
the same paths described above in the quark-quark case. The
indices run from $1$ to $N_c^2-1$, and ${\rm Tr}$ denotes the trace in
the adjoint representation. The physical amplitude can be obtained
from the corresponding Euclidean correlator of Wilson lines
$\CA_E^{gg} $ by means of the same analytic continuation used in the
quark-quark case, Eq.~\eqref{ancontrel}. Analogous crossing-symmetry
relations can be derived along the lines of Ref.~\cite{GM}, which are
obtained by combining the analytic continuation with the relation 
\begin{equation}
  \label{crossrelgg}
\CA_{E\, ab,cd}^{gg}(\pi-\theta,b,T) =
        \CA_{E\, ab,dc}^{gg}(\theta,b,T) \,,   
\end{equation}
which follows from the Euclidean symmetries and the reality of the
adjoint representation. 

Since $(V_i)_{ab} = 2{\rm tr}[W_i^{\dag} t^a
W_i^{\phantom{\dag}} \!t^b]$, the expectation value in
Eq.~\eqref{corrfuncdefgg} can be expressed in terms of fundamental and
anti-fundamental Wilson lines, and so we can compute it through the
AdS/CFT correspondence by making use of the technique described
above. In particular, to extract the ``octet'' component of the
amplitude it suffices to contract it with the appropriate invariant
tensors,
\begin{align}
 &\CA^{gg}_{N_c^2-1,A} = -f^{abm}f^{cdm}\CA^{gg}_{E\,ab,cd}
\cr &\phantom{aaa}=  Z_V^{-2}\bigg\langle 
{\rm tr} W_1^{\dag} {\rm tr} W_2^{\dag} {\rm tr}[W_1 W_2]  
\!-\!{\rm tr} W_1 {\rm tr} W_2^{\dag} {\rm tr}[W_1^{\dag} W_2]
\bigg\rangle\,, \label{ggocteta}\\
&\CA^{gg}_{N_c^2-1,S} = \phantom{-}d^{abm}d^{cdm} \CA^{gg}_{E\,ab,cd}
\\ &\phantom{aaa}=  Z_V^{-2}\bigg\langle 
{\rm tr} W_1^{\dag} {\rm tr} W_2^{\dag} {\rm tr}[W_1 W_2]  
\!+\!{\rm tr} W_1 {\rm tr} W_2^{\dag} {\rm tr}[W_1^{\dag} W_2]
\!-\!{2 \over N_c} 
|{\rm tr} W_1
{\rm tr} W_2|^2
 \bigg\rangle\,, \label{ggoctets} 
 \end{align}
and moreover $Z_V=\langle |\tr W_i|^2 -1\rangle/(N_c^2-1)$. 
By construction, the quantities $\CA^{gg}_{N_c^2-1,S}$ and
$\CA^{gg}_{N_c^2-1,A}$ are respectively even and odd under $\theta\to
\pi-\theta$, thus corresponding to crossing-even and crossing-odd
amplitudes after analytic continuation. As we will see further on,
their evaluation by means of the 
AdS/CFT correspondence and minimal surfaces reduces basically to the
quark-quark case discussed above.

\nsection{Minimal surface for gluon-gluon scattering in the
Alday-Maldacena approach} 

One of the main differences between the two methods described in the
previous section is that the boundaries of the relevant minimal
surfaces are given in the Minkowskian IR region, in the
Alday-Maldacena case, and in the Euclidean UV region, in the
case of the eikonal approach. While there is no contradiction in this,
as we have already explained above, it is nevertheless interesting to
investigate the issue of boundaries, to see if a connection can be
found between the two cases.

In this section we discuss in some detail the geometric structure 
of the minimal surface in anti-de Sitter space found in Ref.~\cite{AM}. 
Firstly, in the next subsection, we recall the behavior 
of the Alday-Maldacena solution in position space, Eq.~\eqref{solorig},
near the IR boundary in ordinary (Minkowskian) $AdS_5$. 
Then, starting from this solution 
and performing an analytic continuation, 
we obtain a related minimal surface in Euclidean $AdS_5$, 
which in a sense defines the near-UV boundary behavior of \eqref{solorig}. 
Finally, we discuss the possible relation between this surface and the
minimal surface relevant to quark-quark scattering in the eikonal approach.

\subsection{The IR boundary}

We shall investigate the near-boundary behavior
of the minimal surface \eqref{solorig}, relevant to gluon-gluon scattering. 
In particular, we shall be interested in the Regge domain
$s /t \gg 1$ and $t$ fixed, which in terms of the surface
parameters $\alpha$ and $\beta$ defined in Eq.~\eqref{mandpara} implies
 $1-\beta = ({\alpha/ \pi})\sqrt{2 / (-s)}\to 0$ and $\alpha \to \pi
 \sqrt{2(-t)}$. 

For later convenience, we rewrite the solution \eqref{solorig} as 
\begin{subequations}
\label{solLC}
  \begin{align}
 x_0 &= -i{R^2 \over 2\alpha}\sqrt{1+\beta^2}\sinh u_+ \sinh u_- \,,
 \label{solLC{a}} \\
x_+ &= -i{R^2 \over 2\alpha} \big[ (1+\beta)u_- +(1-\beta) \cosh u_+
\sinh u_- \big] \,, \label{solLC{b}} \\
x_- &= i{R^2 \over 2\alpha} \big[ (1-\beta)u_+ +(1+\beta) \sinh u_+
\cosh u_- \big] \,, \label{solLC{c}}\\ 
x_3 &= 0 \,, \label{solLC{d}}\\
z &= {R^2 \over 2\alpha}\big[ (1+\beta)\cosh u_+ +(1-\beta)\cosh u_-
\big] \,, \label{solLC{e}}  
  \end{align}
\end{subequations}
where we have redefined the coordinates as
$$
x_\pm \equiv x_1 \pm x_2 \,, \quad u_\pm \equiv u_1 \pm u_2 \,.
$$
Note that the factors $1-\beta$ and $1+\beta$ are proportional 
to the inverse of the square root of the Mandelstam variables, 
$(-s)^{-1/2}$ and $(-t)^{-1/2}$, respectively (see Eq.~\eqref{mandpara}). 
Since Eq.~\eqref{solLC{e}} implies $z \geq R^2/\alpha$, 
the minimal surface described by Eqs.~\eqref{solLC} reaches the IR boundary
$z=\infty$ of AdS, but is bounded apart from the UV boundary, $z=0$.

We analyze now the IR behavior around the boundary $z=\infty$ 
in the complexified $AdS_5$ space. Here we are considering 
the region $0\leq \beta < 1$,  
while the forward Regge limit $\beta = 1$ will be studied later. 

There are four possibilities in order for the minimal surface to reach
the IR boundary $z = \infty$, namely $u_+ = \pm \infty$ or $u_- = \pm
\infty$. We consider first the case $u_+ \to \pm \infty$ at fixed $u_-$.
The solution \eqref{solLC} is then approximated by
\begin{equation}
  \label{irapplus}
\begin{aligned}
&x_0 \approx \mp i{R^2 \over 4\alpha}\sqrt{1+\beta^2} e^{\pm u_+} \,, \quad 
x_+ \approx -i{R^2 \over 4\alpha}(1-\beta)e^{\pm u_+} \sinh u_- \,, \\
&x_- \approx \pm i{R^2 \over 4\alpha}(1+\beta)e^{\pm u_+} \cosh u_- \,, \quad
z \approx {R^2 \over 4\alpha}(1+\beta)e^{u_+} \,.    
  \end{aligned}
\end{equation}
From these equations, we obtain 
\begin{align}
 &x_0 = \pm{\sqrt{1+\beta^2} \over 1-\beta}x_+ \,, \label{irapxzero}\\
&\bigg({z \over 1+\beta}\bigg)^2 = \bigg({x_+ \over
1-\beta}\bigg)^2 -\bigg({x_- \over 1+\beta}\bigg)^2 \,. \label{irapz}  
\end{align}
On the $(x_+,x_-)$-plane with $z = z_{\rm IR}(\gg 1)$ fixed, 
Eq.~\eqref{irapz} defines the hyperbola
\begin{equation}
  \label{projHBplus}
z_{\rm IR}^2 = f_{u_+}(\Im x_+, \Im x_-; \beta) \equiv
-\bigg({1+\beta \over 1-\beta}\bigg)^2 (\Im x_+)^2 + (\Im x_-)^2
\,.   
\end{equation}
Note that Eq.~\eqref{irapplus} implies that $x_0$, $x_+$ and $x_-$ are
purely imaginary.  

We consider now the case $u_- \to \pm \infty$ with $u_+$ fixed.
The solution \eqref{solorig} is approximated by
\begin{equation}
  \label{irapminus}
  \begin{aligned}
&x_0 \approx \mp i{R^2 \over 4\alpha}\sqrt{1+\beta^2} e^{\pm u_-}\sinh u_+ \,, \quad
x_+ \approx \mp i{R^2 \over 4\alpha}(1-\beta)e^{\pm u_-}\cosh u_+ \,, \cr
&x_- \approx i{R^2 \over 4\alpha} (1+\beta)e^{\pm u_-}\sinh u_+ \,, \quad
z \approx {R^2 \over 4\alpha}(1-\beta)e^{\pm u_-} \,.    
  \end{aligned}
\end{equation}
These equations lead to 
\begin{align}
&x_0 = \mp{\sqrt{1+\beta^2} \over 1+\beta}x_- \,, \label{irapxzeroM} \\
&-\bigg({z \over 1-\beta}\bigg)^2 = \bigg({x_+ \over 1-\beta}\bigg)^2 -\bigg({x_- \over 1+\beta}\bigg)^2 \,. \label{irapzM}
\end{align}
Fixing $z = z_{\rm IR}$, Eq.~\eqref{irapzM} defines the hyperbola 
\begin{equation}
  \label{projHBminus}
z_{\rm IR}^2 = f_{u_-}(\Im x_+, \Im x_-; \beta) \equiv (\Im x_+)^2 - \bigg({1-\beta \over 1+\beta}\bigg)^2 (\Im x_-)^2 \,.  
\end{equation}
\begin{figure}[t]
  \centering
  \epsfbox{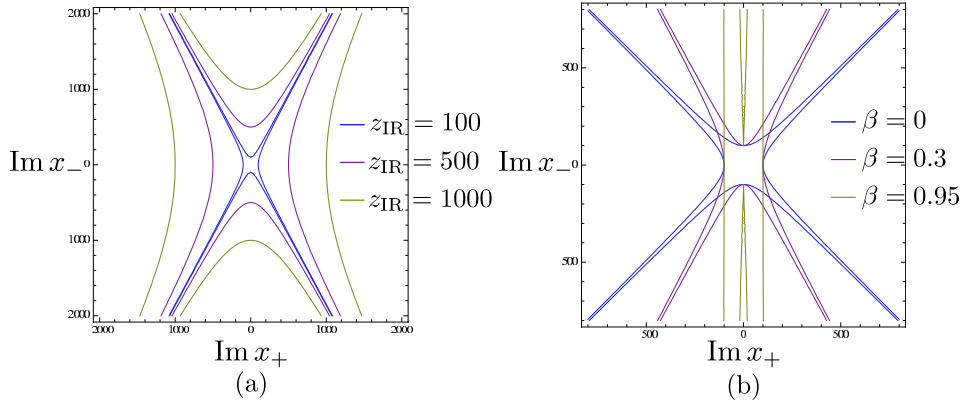}
  \caption{
(a) $z_{\rm IR}^2 = f_{u_\pm}(\Im x_+, \Im
    x_-;0.3)$ with $z_{\rm IR} = 100,500,1000$. \\ 
      (b) $100^2 = f_{u_\pm}(\Im x_+, \Im x_-;\beta)$ with $\beta =
      0,0.3,0.95$.} 
\label{figirzbeta}
\end{figure}
The hyperbolae \eqref{projHBplus} and \eqref{projHBminus} are shown in
Fig.~\ref{figirzbeta}. At fixed $\beta$, the hyperbolae escape to spatial
infinity, {\it i.e.}, in the $(x_+,x_-)$-plane, as $z_{\rm IR} \to \infty$,
see Fig.~\ref{figirzbeta} a. At fixed $z_{\rm IR}$, the angle between the
asymptotes of the hyperbolae tends to zero as $\beta\to 1$, 
see Fig.~\ref{figirzbeta} b. 
\begin{figure}[t]
  \centering
  \epsfbox{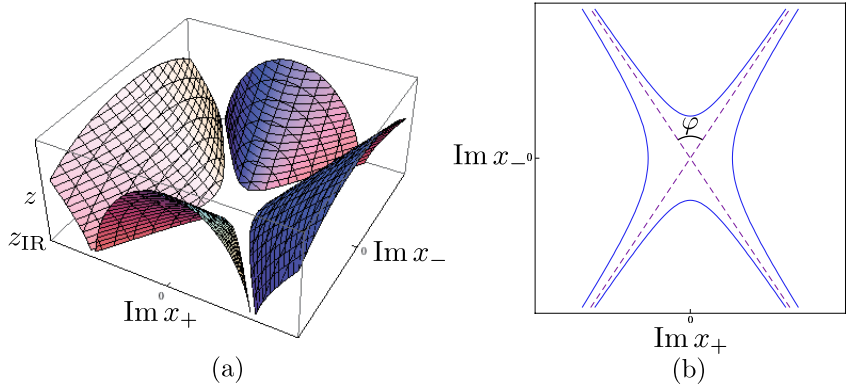}
  \caption{
(a) The minimal surface determined by
    Eqs.~\eqref{irapz} and \eqref{irapzM}. \\
      (b) The behavior of the surface at fixed $z_{\rm IR} (\geq
      R^2/\alpha)$.
}
\label{figirbdry}
\end{figure}
To further clarify the IR behavior of the minimal surface, we show in
Fig.~\ref{figirbdry} a the plot of Eqs.~\eqref{irapz} and \eqref{irapzM}. 
The surface blows up and escapes to spatial infinity when $z$ becomes larger. 
In Fig.~\ref{figirbdry} b, we show again the hyperbolae defined 
in Eqs.~\eqref{projHBplus} and \eqref{projHBminus}, together with
their asymptotes.  
The physical scattering angle $\varphi$ in the $u$-channel is given by
\begin{equation}
  \label{scatangle}
\tan {\varphi \over 2} = \sqrt{t \over s} = {1- \beta \over 1 + \beta}   \,,  
\end{equation}
and so it is equal to the angle formed by the asymptotes. 
Comparison of Fig.~\ref{figirbdry} b with Fig.~\ref{figirzbeta} b then
shows clearly that the scattering angle goes to zero when $\beta \to
1$, that is, in the Regge limit. 

In principle, the momentum space formulation of the minimal surface
problem considered by Alday and Maldacena can be traded for a
coordinate space formulation, with the more complicated boundary
discussed above. This is closer in spirit to the variational problem
encountered in the eikonal approach, although the boundary in the two
cases are living in spaces with different signature, and still on
opposite ends of AdS.

\subsection{The UV boundary: analytic continuation to Euclidean AdS}

The minimal surface solution \eqref{solorig} lives 
in the complexified anti-de Sitter space. If we now perform the
following analytic continuation of the world-sheet coordinates,  
$$
u_\pm = i w_\pm \,, 
$$
the coordinates $x_\pm$ become real for real $w_\pm$. Since $x_0$ is
still complex, we perform additionally the Wick rotation $x_4 =
ix_0$. We then obtain a new minimal surface, given by 
\begin{subequations}
\label{solEuc}
\begin{align}
x_+ &= {R^2 \over 2\alpha}\bigl[ (1+\beta)w_- +(1-\beta)\cos w_+ \sin
w_-\bigr] \,, \label{solEuc{a}} \\ 
x_- &= -{R^2 \over 2\alpha}\bigl[ (1-\beta)w_+ +(1+\beta)\sin w_+ \cos
w_-\bigr] \,, \label{solEuc{b}} \\
x_3 &= 0 \,, \label{solEuc{c}} \\
x_4 &= -{R^2 \over 2\alpha}\sqrt{1+\beta^2}\sin w_+ \sin w_-
\,, \label{solEuc{d}} \\
z &= {R^2 \over 2\alpha}\bigl[(1+\beta)\cos w_+ +(1-\beta)\cos
w_-\bigr] \,, \label{solEuc{e}} 
\end{align}
\end{subequations}
in the real Euclidean anti-de Sitter space. 
\begin{figure}[t]
  \centering
\epsfbox{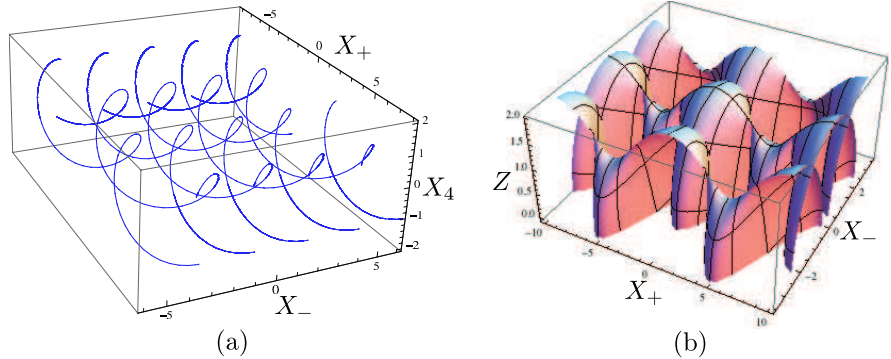}
\caption{(a) The multiple helices as the UV boundary of Eqs.~\eqref{solEuc}
  at $z=0$ with $\beta = 0.6$. \\ 
   (b) The minimal surface \eqref{solEuc} with $\beta = 0.6$.}
\label{fighelices}
\end{figure}
Eq.~\eqref{solEuc{e}} implies that the minimal surface reaches the UV
boundary $z=0$, where it describes a multiple helices configuration
(see Fig.~\ref{fighelices} a). The axes in Fig.~\ref{fighelices}
correspond to the coordinates $X_\pm, X_4$ and $Z$, defined by the
rescaling $X_{\pm,4} = (2\alpha / R^2) x_{\pm,4}$ and $Z = (2\alpha /
R^2) z$. The minimal surface \eqref{solEuc} is depicted in
Fig.~\ref{fighelices} b. This construction provides 
a Euclidean formulation of the Alday-Maldacena minimal surface, with
boundaries in the UV region, which can be directly compared with the
minimal surface problem relevant to the eikonal approach.

A comment is in order here. In Refs.~\cite{DJW,DDJK}, a family of
classical string solutions in $AdS_3 \times S^3$ was discussed in
terms of the Pohlmeyer reduction of the string sigma
model. Ref.~\cite{DDJK} obtained a space-like surface in $AdS_3$ 
with conformal complex world-sheet coordinates and embedded it into
$AdS_5$, so that the Alday-Maldacena type solution\footnote{This solution
has a rotated version of the boundary condition of Ref.~\cite{AM}.} was
reproduced. Then, by Wick rotation of the world-sheet time coordinate,
Ref.~\cite{DDJK} found time-like surfaces in $AdS_3$, one of which had
helicoid geometry. This surface is similar to the one with the double
helix boundary that we obtain in the limit $\beta\to 1$, discussed
below; however, our Wick rotation and analytic continuations are
different from those of Refs.~\cite{DJW,DDJK}.

\subsection{The forward Regge limit of the UV boundary}

We consider now the forward Regge limit, 
\begin{equation}
  \label{Reggelimit}
-s \to \infty,\quad -t \ {\rm fixed},  
\end{equation}
of the solution \eqref{solEuc}. In this limit the Mandelstam variable $u$
goes to $+\infty$, because of the relation $s+t+u=0$. Using the
relation~\eqref{mandpara} between the parameters $\alpha, \beta$ of the
minimal surface and the Mandelstam variables $s, t$, the
limit~\eqref{Reggelimit} is seen to correspond to $\beta=1$.
Since in this limit the scattering angle vanishes, $\varphi= 0$, 
we are dealing here with forward Regge scattering.\footnote{Note that 
the value of $\alpha$ and thus
of $-t=\alpha^2/(2\pi^2)$ is arbitrary, but fixed, in this forward
Regge limit.} 

In the forward Regge limit, 
the minimal surface \eqref{solEuc} in Euclidean space is reduced to 
\begin{equation}
  \label{extRegge}
  \begin{aligned}
&x_+ = {R^2 \over \alpha}w_- \,, \quad 
x_- = -{R^2 \over \alpha}\sin w_+ \cos w_- \,, \\
&x_4 = -{R^2 \over \sqrt{2} \alpha}\sin w_+ \sin w_- \,, \quad 
z = {R^2 \over \alpha}\cos w_+ \,, 
  \end{aligned}
\end{equation}
and $x_3 = 0$. At the UV boundary $z=0$, this surface describes a
double helix,
\begin{figure}[t]
  \centering
  \epsfbox{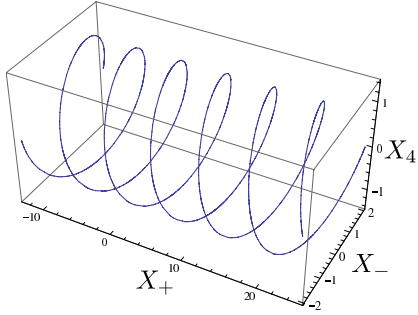}
  \caption{The double helix forming the UV boundary in the forward Regge limit.}
  \label{figdspiral}
\end{figure}
\begin{equation}
  \label{helicRegge}
x_+ = {R^2 \over \alpha}w_- \,, \quad 
x_- = \pm{R^2 \over \alpha}\cos w_- \,, \quad
x_4 = \pm{R^2 \over \sqrt{2}\alpha}\sin w_- \,.  
\end{equation}
The double helix \eqref{helicRegge}, depicted in Fig.~\ref{figdspiral}, 
is reminiscent of the boundary of the minimal surface 
that was used in Ref.~\cite{JPi} in the computation 
of the quark-quark scattering amplitude in the eikonal approach. 
We shall comment on this in the following subsection. 

\subsection{Relation with the minimal surface for quark-quark scattering
in the eikonal approach} 

In the previous subsection, we have obtained 
the double helix \eqref{helicRegge} (see Fig.~\ref{figdspiral}) 
as the boundary of the Euclidean minimal surface \eqref{extRegge}, 
which appears in the forward Regge limit for gluon-gluon scattering. 
The boundary of this surface lies on the UV boundary 
of (Euclidean) anti-de Sitter space. 
On the other hand, the double helix appears in the context of
quark-quark scattering in the eikonal approximation
\cite{JP,JPi,JPii}, as the IR cutoff of a truncated ``generalized
helicoid''.  
Indeed, as we have recalled, the minimal surface relevant
to quark-quark scattering, 
defined by the straight line boundaries \eqref{lines}, 
was studied in Ref.~\cite{JPi} by making the ``generalized helicoid'' 
ans{\" a}tz \eqref{helico}. When truncating the surface in order to
regularize its area, as in Eq.~\eqref{JParea}, 
the double helix appears in the projection of the surface on the UV boundary. 

Is there a relation between the minimal surfaces used in the
Alday-Maldacena approach and in the eikonal approach? 
\begin{figure}[t]
  \centering
  \epsfbox{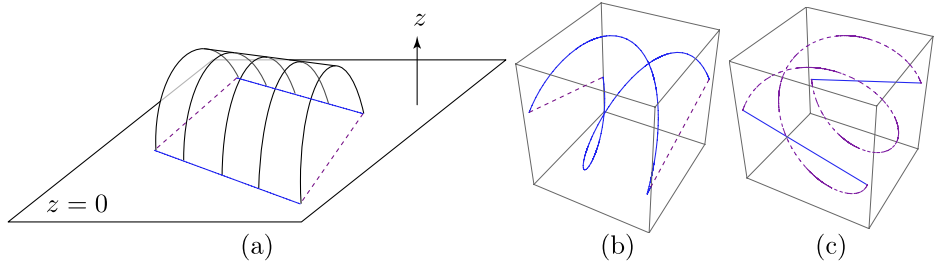}
  \caption{(a) The minimal surface with parallel line
segments as the boundary. \\
  (b), (c) The UV boundary and cutoff in the $(x_+,x_-,x_4)$ space at
 $z=0$.}
\label{figwilsonloop}
\end{figure}
One can intuitively represent the situation as in Fig.~\ref{figwilsonloop}. 
Let us imagine first the minimal surface with two parallel
straight-line segments as the UV boundary at $z=0$ in (Euclidean)
anti-de Sitter space (Fig.~\ref{figwilsonloop} a). This corresponds to
the well-known 
configuration of two parallel Wilson lines used for the computation of
the quark-quark potential. The solid line segments in Fig.~\ref{figwilsonloop} a
describe the boundaries of the minimal surface at $z=0$, 
while the dotted lines are defined by the IR cutoff imposed on the surface. 
Twisting the dotted line segments in the $(x_+,x_-,x_4)$ space, 
we obtain the double helix (Fig.~\ref{figwilsonloop} b). 
This is exactly the geometry of Eqs.~\eqref{helicRegge}, that is obtained 
in the forward Regge limit of gluon-gluon scattering in the
Alday-Maldacena approach. 
On the other hand, by twisting the solid line segments in
Fig.~\ref{figwilsonloop} a,  
we obtain Fig.~\ref{figwilsonloop} c, in which the dotted line segments become
the double helix. This is the configuration that is desired in
computing the quark-quark scattering amplitude in the eikonal
approach. The solid lines describe the trajectories of quarks and the
dotted lines are determined by the IR cutoff.

The full answer to the question raised above requires the exact
analytic determination of the minimal surface having the boundary
configuration of Fig.~\ref{figwilsonloop} c, relevant to quark-quark scattering,
which could then be compared to the minimal surface found in Ref.~\cite{AM}
for gluon-gluon scattering. However, the exact solution to this
problem has not been found yet. Nevertheless, as we will see further
on in Section 5,  
new insights can be obtained by performing a convenient conformal
transformation on the minimal surface, and by critically reconsidering
the study of the ``generalized helicoid'' ans{\" a}tz \eqref{helico}.

\nsection{Regge behavior of scattering amplitudes
in the \\ Alday-Maldacena approach}

In this Section we discuss the behavior of the gluon-gluon scattering
amplitude \eqref{ampAM}, and of the quark-quark scattering amplitude of
Ref.~\cite{BV}, in the Regge limit $-s\to\infty$, $t$ fixed (see
also Ref.~\cite{NS}), both in the momentum representation and in the
impact-parameter representation.

\subsection{Momentum representation}

In order to display the Regge behavior of the four-gluon scattering
amplitude Eq.~\eqref{ampAM}, it is convenient to expand the divergent
contributions \eqref{ampAMdiv} with respect to $\epsilon$. 
One then obtains
\begin{equation}
  \label{Sdiv}
iS_{\rm div}(p) = -{1 \over \epsilon^2}{\sqrt{\lambda} \over 2\pi}
	+{1 \over \epsilon}{\sqrt{\lambda} \over 4\pi}\biggl( \log{-p \over \mu^2}
	-1+\log 2 \biggr) -{f(\lambda) \over 16}\biggl(\log{-p \over \mu^2}\biggr)^2 
	+{g(\lambda) \over 8}\log{-p \over \mu^2} +\CO(\epsilon)\,,  
\end{equation}
where $p=s,t$, and where 
we have denoted 
\begin{equation}
  \label{cuspanomdim}
 f(\lambda) = {\sqrt{\lambda} \over \pi} \,, \quad 
g(\lambda) = {\sqrt{\lambda} \over \pi}(1-\log 2) \,.
\end{equation}
The meaning of $f(\lambda)$ and $g(\lambda)$ becomes clear if we
rewrite Eq.~\eqref{Sdiv} in terms of a new IR cutoff $m$, defined
as\footnote{Since $\epsilon$ is negative, $\epsilon\to 0_-$ corresponds to
$m/\mu\to 0$, {\it i.e.}, to an IR cutoff.}
\begin{equation}
  \label{uvparam}
{1 \over \epsilon} \equiv \log {m \over \mu} \,.  
\end{equation}
Neglecting terms which do not depend on $p$, we obtain
\begin{equation}
  \label{Sdivbis}
iS_{\rm div}(p) = - {f(\lambda) \over 16}\biggl(\log{-p \over m^2}\biggr)^2
+{g(\lambda) \over 8}\log{-p \over m^2} + (p\hbox{-independent terms})\,,  
\end{equation}
with $f(\lambda)$ appearing in front of the leading IR-divergent term
proportional to $(\log m)^2$, and $g(\lambda)$ appearing in front of
the subleading ($\log m$) divergence. 

It is important to note that $f(\lambda)$ appears in the expression of 
the cusp anomalous dimension $\Gamma_{\rm cusp}(\gamma)$, which
represents the contribution of a cusp of boost parameter $\gamma$ to
the vacuum expectation value of a  Wilson loop in the fundamental
representation. 
For large $|\gamma|$, one has indeed 
$\Gamma_{\rm cusp}(\gamma)\simeq -\bigl(f(\lambda)/4\bigr)|\gamma|$. 
The cusp anomalous dimension \cite{Po,Ko} is relevant also 
for the calculation of the anomalous dimension $\gamma_S$ 
of twist-two operators of large spin $S$, 
$\gamma_S \simeq f(\lambda)\log S$ 
(see, {\it e.g.}, Ref.~\cite{Kr} and references therein).  

Using the expansion \eqref{Sdiv} and the definitions
\eqref{cuspanomdim} and \eqref{uvparam},
the expression of the amplitude \eqref{ampAM} simplifies 
\begin{align}
\CA^{\rm gluon} (s,t)&= C_\epsilon \biggl({-s \over m^2}\biggr)^{-{f(\lambda)
\over 4} \log {-t \over m^2}+{g(\lambda) \over 4}} 
	\biggl({-t \over m^2}\biggr)^{g(\lambda) \over 4} \,,
        \label{amptotal} \\
C_\epsilon &= \exp\biggl(-{\sqrt{\lambda} \over \pi}{1 \over
\epsilon^2} +{\tilde C} +\CO(\epsilon)\biggr) \,. \label{ampepsilon} 
\end{align}
We note that the terms $\log(-s/\mu)^2$ and $\log (-t/\mu)^2$ in the
finite part of Eq.~\eqref{ampAM} are compensated by corresponding terms of
order $\epsilon^0$ coming from the expansion \eqref{Sdiv} of $S_{\rm
  div}$~\cite{NS}.

It is important to realize that formula \eqref{amptotal} has precisely the
form of a Regge amplitude~\cite{NS,DKS} (in particular, it is
Regge-exact in the $s$-channel~\cite{DKS}). Indeed, including for
completeness also the Born term factor, which for large $-s$ and fixed
$t$ reads  
\begin{equation}
  \label{tree}
\CA_{\rm tree} \propto {-s \over -t} \,,  
\end{equation}
the gluon-gluon scattering amplitude is of the form
\begin{equation}
  \label{amptotRegge}
\CA (s,t) = \CA_{\rm tree}\,\CA^{\rm gluon} (s,t) = \beta(t)\biggl({-s
\over m^2}\biggr)^{\alpha(t)},   
\end{equation}
where $\alpha(t)$ is the {\it Regge trajectory},
\begin{equation}
  \label{trajectory}
  \begin{aligned}
&\alpha(t) = \alpha_0(t) + \alpha_1 \,, \\
&\alpha_0(t) = -{f(\lambda) \over 4}\log{-t \over m^2} \,, \quad
\alpha_1 = {g(\lambda)\over 4} +1 \,,    
  \end{aligned}
\end{equation}
and where $\beta(t)$ is given by
\begin{equation}
  \label{residue}
\beta(t) \propto  C_\epsilon\biggl({-t \over m^2}\biggr)^{{g(\lambda)
\over 4}-1}\,,   
\end{equation}
up to a $t$-independent constant. In the large-$N_c$ limit, the
dominant contribution to the amplitude comes from the trajectory with
the quantum numbers of the gluon (see Ref.~\cite{DKS} and references
therein), so that $\alpha(t)$ is identified as the gluon Regge
trajectory. 

In the expression of the amplitude \eqref{amptotal}, 
one may further distinguish the separately factorized terms 
in $s$ and $t$ from the non-factorizable one, namely 
\begin{align}
&\CA^{\rm gluon} (s,t) = C_\epsilon \CA_{\rm fact}(s) \CA_{\rm
fact}(t) \CA_{\rm nonfact}(s,t)\,, \label{ampconv} \\
&\CA_{\rm fact}(p) = \exp \biggl({g(\lambda) \over 4} \log{-p \over
m^2}\biggr) \,, \quad (p=s,t) \label{ampfact} \\
&\CA_{\rm nonfact}(s,t) = \exp \biggl(-{f(\lambda) \over 4}\log{-s
\over m^2} \log{-t \over m^2}\biggr)  \,. \label{ampregge}
\end{align}
As it is well known, the non-factorizable expression \eqref{ampregge}
characterizes the $t$-dependence of the leading Regge trajectory for
``octet'' $t$-channel exchange, 
$\alpha_0(t)$ in Eqs.~\eqref{trajectory}. 
This term is independent of
the particular choice of the IR cutoff: indeed, a rescaling of the IR
cutoff  $m\to e^\kappa m$ leaves it unchanged. On the other
hand, the same rescaling changes the coefficient of the logarithm in
Eq.~\eqref{ampfact}, $g(\lambda) \to g_\kappa(\lambda)= g(\lambda)  + 2\kappa
f(\lambda)$, as well as the constant $C_\epsilon \to
C_{\epsilon,\kappa} = C_\epsilon e^{-\kappa^2 f(\lambda)}$. This
results in the dependence of the factorizable terms of the amplitude
\eqref{amptotRegge} on the regularization scheme: this is not surprising,
given the regularization-scheme dependence of the gluon Regge trajectory. 
Indeed, a calculation in the radial-cutoff scheme, $i.e.$, limiting the
integration of the area of the minimal surface \eqref{solTdual} to $r>r_c$,
gives~\cite{AMr} 
\begin{equation}
  \label{AMrad}
  \CA_{\rm radial}^{\rm gluon}(s,t) = \exp\biggl(-{f(\lambda)\over 4} \log
  {-s \over \tilde m^2}\log {-t \over \tilde m^2} + {\rm
    const.}\biggr)\,, 
\end{equation}
where $\tilde m = r_c/(2\sqrt{2}\pi)$, corresponding to a gluon Regge
trajectory with $\alpha_1=1$. It is easy to see that the
$s,t$-dependent terms in the two schemes are related by the rescaling
$\tilde m=m\sqrt{e/2}$ of the IR cutoffs.

The key property expected for a Regge trajectory is to be ``universal'',
{\it i.e.}, present in all high-energy channels at fixed momentum transfer 
for the same exchanged quantum numbers. 
This leads us to compare the results for gluon-gluon scattering
discussed above, especially the Regge trajectory \eqref{trajectory},  
with the quark-quark elastic scattering amplitude obtained in
Ref.~\cite{BV}, along the lines of the Alday-Maldacena approach. 
We report here only the final result for
the color-independent part of the amplitude (divided by the tree
amplitude) obtained in the limit of small quark masses, which reads
\begin{equation}
  \label{qqBV}
  \CA^{\rm quark} = \exp\bigg[ -{f(\lambda) \over 4} \log
  {-t \over \tilde m^2}\left(\log {-s \over m_1 m_2}
    -2(\log(\sqrt{2}-1)+1)\right) + {\rm const.}\bigg] \,,
\end{equation}
where $\tilde m = r_c/(2\sqrt{2}\pi)$, with
$r_c$ the radial cutoff used in the calculation,  and $m_{1,2}$ are
the quark masses. The result above holds as long as $1 \ll -s/(m_1m_2)
\ll r_c^{-1}$, which implies that one cannot take the large-$s$ limit
at fixed cutoff. Nevertheless, the term $ \log (-s)\log (-t)$ is 
not affected by a change of the cutoff, which implies that it is
reliably captured by the approximation. On the contrary, this is not
true for  $\log (-s)$ and $\log (-t)$ terms, which are therefore not
completely under control at the present stage. It is immediate to see
that also this amplitude is of Regge type, with the same $t$-dependent
part for the Regge trajectory as in the gluon-gluon case, as expected
from universality.

\subsection{Impact-parameter representation}

For further comparison with the eikonal approach,
we derive now the impact-parameter representation
for the gluon-gluon scattering amplitude. 
The impact-parameter amplitude ${\tilde \CA}^{\rm gluon}({\hat \chi},b)$ 
is obtained by performing the two-dimensional 
Fourier transform of the amplitude $\CA(s,t)$ with respect to the
transverse momentum. 
Setting $-t = k^2$ with $k$ the modulus of the transverse momentum, 
and including the usual factor $s^{-1}$ in the definition 
of the impact-parameter amplitude, 
we obtain at large $-s$ (up to an irrelevant constant)
\begin{equation}
  \label{fourier}
{\tilde \CA}^{\rm gluon}({\hat \chi},b) = C_\epsilon \int {dk \over k}\, J_0(kb) 
\CA^{\rm gluon}(s,t=-k^2) \,,   
\end{equation}
where the hyperbolic angle ${\hat \chi}$ is defined as 
\begin{equation}
  \label{Gluonchi}
{\hat \chi} = \log {-s  \over  m^2},   
\end{equation}
as appropriate for a $u$-channel process. 
Azimuthal invariance has been taken into account to reduce the
two-dimensional Fourier transform to a Hankel transform of order 0,
involving the ordinary Bessel function $J_0(\zeta)$ with $\zeta=kb$. 

Inserting the amplitude \eqref{ampconv} into Eq.~\eqref{fourier}, one
obtains
\begin{align}
  {\tilde \CA}^{\rm gluon}({\hat \chi},b) &= C_\epsilon \bigl( m^2 b^2 \bigr)^{-{1 \over 4}h({\hat \chi};\lambda)} 
	e^{{g(\lambda) \over 4}{\hat \chi}} K({\hat \chi})  \,,  \label{ampRegge}\\
h({\hat \chi};\lambda) &\equiv -f(\lambda) {\hat \chi} +g(\lambda) \,,
\nonumber  
\end{align}
where
\begin{equation}
  \label{Gchi}
K({\hat \chi}) \equiv \int_0^\infty d\zeta\, \zeta^{{1 \over
2}h({\hat \chi};\lambda)-1} J_0(\zeta)  
	= 2^{{h \over 2}-1} {\Gamma \bigl({h \over 4}\bigr)\over
        \Gamma\bigl(1-{h \over 4}\bigr)} \,.   
\end{equation}
The integral~\eqref{Gchi} is convergent in a limited parametric region for
$h({\hat \chi};\lambda)$, namely $0 < h < 3$, which lies away from the
physical Minkowski region where ${\hat \chi} \gg 1$, that is, $h \ll 0$.  
This is due to the form of the amplitude \eqref{ampregge},  which for
$h({\hat \chi};\lambda)$ outside of the above-mentioned domain makes
the integrand of Eq.~\eqref{fourier} too singular at small $k$. 
We can however reach the physically interesting region by means of
analytic continuation\footnote{The analytic continuation is made  
passing from $h>0$ to $h<0$ in the lower half of the complex plane,
{\it i.e.}, $h \to |h| e^{-i(\pi - \delta)}$, 
in order to avoid the poles of the Gamma function on the real negative axis. 
This choice is consistent with the usual ``$-i\epsilon$'' prescription,
{\it i.e.}, $m^2\to m^2-i\epsilon$, which in the case at hand implies that
${\hat \chi}$ acquires a small positive imaginary component. After using the
Stirling approximation at large $|h|$, one takes the limit $\delta \to
0$.
}  
of the function $K({\hat \chi})$ defined in Eq.~\eqref{Gchi}, which in
the high-energy 
Minkowski region where ${\hat \chi} \gg 1$ becomes
\begin{equation}
  \label{Gchianal}
K({\hat \chi}) \approx {1 \over e}\bigg({2e \over -h}\bigg)^{-{h \over 2} +1}
		\exp\bigg[i\pi \bigg({1 \over 2} -{h \over
                  4}\bigg)\bigg] \,,    
\end{equation}
where we have made use of Stirling's formula, $\Gamma(z) \sim
\sqrt{2\pi} e^{-z}z^{z-{1 \over 2}}$ (for $|z| \to \infty$).  
Since $h \approx -f(\lambda) {\hat \chi}$ for ${\hat \chi} \gg 1$, in
the Minkowski region, we may write the following expansion in energy
\begin{equation}
  \label{Glogchianal}
\log K({\hat \chi})=  -{f(\lambda) \over 2}{\hat \chi}
	\bigg(\log {\hat \chi} +\log {f(\lambda) \over 2e} -i{\pi
          \over 2}\bigg)  
	+ \bigg({g(\lambda) \over 2}-1\bigg)\log {\hat \chi} +\cdots \,, 
\end{equation}
where the terms behaving at most as a constant are neglected.

Taking into account the expansion \eqref{Glogchianal}, 
the resulting impact-parameter amplitude \eqref{ampRegge} can then be
rewritten at high energy and in log form as the expansion
\begin{equation}
  \label{Logampl}
  \begin{aligned}
    -\log {\tilde \CA}^{\rm gluon}({\hat \chi},b) &= 
    -{f(\lambda)\over 2}{\hat \chi} \log {mb} + {f(\lambda)\over
      2}{\hat \chi} \log {\hat \chi} 
    +{\hat \chi}\biggl[{f(\lambda) \over 2} \biggl(\log
    {f(\lambda) \over 2e} -i{\pi \over 2}\biggr)  
    -{g(\lambda) \over 4}\biggr] \\
    &\quad + \log{\hat \chi}\biggl(1-{g(\lambda) \over 2}\biggr) 	+
    {g(\lambda) \over 2}\log {mb} +\cdots \,, 
  \end{aligned}
\end{equation}
where the overall sign has been chosen for further comparison with the
minimal area obtained from the eikonal approach\footnote{Note that 
we did not obtain formula \eqref{Logampl} as the area of a minimal surface 
in Euclidean impact-parameter space. 
It may be worth mentioning, nevertheless, 
that it would be interesting to investigate 
if it can be obtained as the solution 
of a properly formulated minimal surface problem in impact-parameter space. 
} 
in the following section.

The result \eqref{Logampl} calls for comments:
\begin{itemize}
\item[i)] The expansion \eqref{Logampl} reflects the fact that the amplitude
\eqref{ampRegge} is the product  
of a non-factorizable function of the two kinematic variables, ${\hat \chi}$
and $b$, times a factorizable term, namely
\begin{equation}
  \label{bchifact}
{\tilde \CA}^{\rm gluon}({\hat \chi},b) = C_\epsilon (m^2 b^2)^{{1 \over 4} f(\lambda)
{\hat \chi}}F({\hat \chi}) B(b) \,,   
\end{equation}
where the factorizable sector $F({\hat \chi})B(b)$ is given by
\begin{equation}
  \label{Fchifact} 
F({\hat \chi}) = e^{{1 \over 4}g(\lambda){\hat \chi}}K({\hat \chi})
\,, 
\quad B(b) = (m^2 b^2)^{-{1 \over 4}g(\lambda)}  \,. 
\end{equation}
The first (non-factorizable) term in \eqref{Logampl} is the origin of the
non-factorizable term in Eq.~\eqref{ampregge}, and thus of the $t$-dependent
part of the Regge trajectory. The role of the second ($s$-dependent
factorizable) term is more subtle, and it is better understood when
going back from impact-parameter to momentum space. When taking the
inverse Fourier transform, the non-factorizable $b$-dependent term
gives rise to a factor 
$\exp[(f(\lambda)/2){\hat \chi} \log{\hat \chi}] 
\sim (\log (-s) )^{\log (-s)}$, which is
not of Regge type and would be the leading dependence on energy, but
which is precisely canceled by the second term.\footnote{Note that on the  
other hand a factor $(\log (-s))^{\rm const.}$ is compatible with a Regge
amplitude, indicating the presence of a multiple pole or of a Regge
cut in the complex-angular-momentum representation of the amplitude. 
} 
These two terms combine into the expression $-(f(\lambda) / 2){\hat
\chi} \log (mb/{\hat \chi})$, which basically encodes the Regge
nature of the amplitude. The third and fourth terms yield a
factorizable $s$-dependence which modifies the Regge trajectory by a
$t$-independent term, and the last term affects the factorizable
$t$-dependent part of the amplitude. 
\item[ii)] The power of $b$  
in Eq.~\eqref{ampRegge} is 
{\it negative} in the convergence region where $0 < h({\hat
  \chi};\lambda) < 3$,  
while it is {\it positive} in the Regge domain $-h({\hat \chi};\lambda)=
f(\lambda) {\hat \chi} -g(\lambda) \gg 0$.  
This is the counterpart in impact-parameter space of the divergence at small
values of $k$ in the Fourier transform \eqref{fourier}. 
Hence an analytic continuation is required 
to obtain the impact-parameter amplitude in the interesting high-energy region.
\item[iii)] The non-factorizable sector in Eq.~\eqref{bchifact}
  depends only on   
the cusp anomalous dimension at high energy, namely
\begin{equation}
  \label{cuspmink}
\Gamma_{\rm cusp}({\hat \chi})
\to  -{f(\lambda) \over 4}{\hat \chi} 
	=  -{\sqrt{\lambda} \over 4\pi}{\hat \chi} \quad 
	{\rm for}\ {\hat \chi} \gg 1 \,.
\end{equation}
It is thus interesting to note that 
the expression \eqref{Logampl} can be rewritten as 
\begin{equation}
  \label{amplcusp}
-\log {\tilde \CA}^{\rm gluon}({\hat \chi},b) 
	\approx 2{\Gamma_{\rm cusp}({\hat \chi})}\ \log {m b \over
          {\hat \chi}} + \cdots \,, 
\end{equation}
where we have neglected terms which are subleading in energy, and 
where we have used the known behavior \eqref{cuspmink} of the cusp anomaly 
for a fundamental Wilson loop in the large-${\hat \chi}$ region.
\item[iv)] It is straightforward to obtain the impact-parameter
representation for the quark-quark scattering amplitude of Ref.~\cite{BV},
Eq.~\eqref{qqBV}, which can be written as
\begin{equation}
  \label{amplcuspqqbv}
-\log {\tilde \CA}^{\rm quark}({\tilde \chi},b) 
	\approx 2{\Gamma_{\rm cusp}({\tilde \chi})}\ \log {\tilde m b
        \over {\tilde \chi}} + \cdots \,,   
\end{equation}
where now ${\tilde \chi}=\log (-s/(m_1 m_2))$ is the hyperbolic angle
between the classical trajectories of the quarks at high energy, and 
the dots stand for ${\cal O}(\tilde\chi)$ terms.
\end{itemize}

\nsection{Quark-quark scattering amplitude in the eikonal approach}

In this section we discuss the minimal surface problem relevant to
quark-quark scattering in the eikonal approach, both from a general
point of view, and exploiting the ``generalized helicoid'' ans{\"
a}tz~\eqref{helico}.

\subsection{General features of the minimal surface}

On general grounds, the area $A^{\rm quark}_{\rm min}$ of the
surface minimizing the functional \eqref{areafunc} has to take the form  
\begin{equation}
  \label{resultJP}
A^{\rm quark}_{\rm min}(\theta,b,T) = \Phi_E(b/T,\theta)
+\Psi_E(\theta) \,,   
\end{equation}
where the splitting between a
$b$-dependent function $\Phi$ and a $b$-independent one $\Psi$ is made
for future convenience.  
This is a consequence of conformal invariance
together with the fact that 
the IR cutoff $T$ is the only length scale other than $b$ that can appear,
once that UV divergencies have been removed.\footnote{This is different,
although similar in spirit, to the argument of Ref.~\cite{JPi}, 
where the UV cutoff $M_B$ appears instead of $1/T$. 
However, as we have explained in Section 2, 
UV divergencies should be absent from the final result.
} 
In particular, as we will see below, the separation between the $\Phi$
and $\Psi$ functions amounts to the product of non-factorizable and
factorizable contributions to the (Euclidean) impact-parameter amplitude.
For future utility, we define the analytic continuation 
of \eqref{resultJP} to Minkowski space as 
\begin{equation}
  \label{resultJPM}
A^{{\rm quark},\,s}_{{\rm min},\,M}(\chi,b,T) = 
A^{{\rm quark}}_{\rm min}(-i\chi,b,iT)\,.
\end{equation}
This quantity enters the $s$-channel
quark-quark scattering amplitude which,
in the minimal surface approximation of the AdS/CFT correspondence,  
is given in impact-parameter space by 
$\tilde \CA^{{\rm quark},\,s}(\chi,b,T) \equiv 
\tilde \CA_E^{{\rm quark}}(-i\chi,b,iT)
= \exp\bigl[-A^{{\rm quark},\,s}_{{\rm min},\,M}(\chi,b,T) \bigr]$, see
Eqs.~\eqref{ancontrel} and \eqref{amplisurf}.
Note that we used the superscript $s$ in the
notations in order to specify the physical channel $s \gg 0$ that we 
consider in Minkowski space.  

Further insight on the structure of $A^{\rm quark}_{\rm min}$ can
be obtained by performing a particular conformal transformation in
Euclidean $AdS_5$ space, namely the inversion of coordinates. 
Such a transformation leaves the area of the surface invariant 
up to a function of the coupling $\lambda$ only~\cite{DG}, 
which is not relevant for our purposes.\footnote{Although the argument 
of Ref.~\cite{DG} is valid for smooth contours, 
we expect that this result holds also for loops with a cusp,
which can be obtained as appropriate limits of smooth loops. 
} 
We can therefore investigate the quark-quark scattering amplitude 
by studying the new minimal surface problem in the inverted coordinates. 

Under the transformation of the target space coordinates
$(x_1,x_2,x_3,x_4,z)$ defined by  
\begin{equation}
  \label{lineTOcircle}
 x_\mu \to x'_\mu = {x_\mu \over |x_\mu|^2 +z^2} \,, \quad 
z\to z' = {z \over |x_\mu|^2 +z^2} \,,
\end{equation}
the Euclidean $AdS_5$ metric is invariant, while the two straight
lines $L_{1,2}$, \eqref{lines}, which define the boundary condition at
$z=0$ in the original coordinates (see Fig.~\ref{figquarkpara}), are
mapped into two circles $C_{1,2}$, which define the boundary at $z'=0$ 
in the new coordinates: 
\begin{equation}
  \label{circles}
  \begin{aligned}
    &C_1: \biggl(-{\sin(\theta/2) \over b}\sin\xi, -{1 \over
b}(1+\cos\xi), 0, {\cos(\theta/2) \over b}\sin\xi \biggr) \,, \cr 
&C_2: \biggl({\sin(\theta/2) \over b}\sin\xi, {1 \over b}(1+\cos\xi),
0, {\cos(\theta/2) \over b}\sin\xi \biggr) \,,  
  \end{aligned}
\end{equation}
where 
\begin{equation}
  \label{circangle}
\sin\xi = {b\tau \over \tau^2 + b^2/4} \,, \quad
	\cos\xi = {-\tau^2 +b^2/4 \over \tau^2 +b^2/4} \,.
\end{equation}
Also in this case, we consider the variational problem for 
$\tau \in [-\infty,\infty]$, {\it i.e.}, for two complete circles, 
and we regularize the area by limiting the integration to $\tau\in [-T,T]$. 
\begin{figure}[t]
  \centering
{\epsfxsize=13.5cm \epsfbox{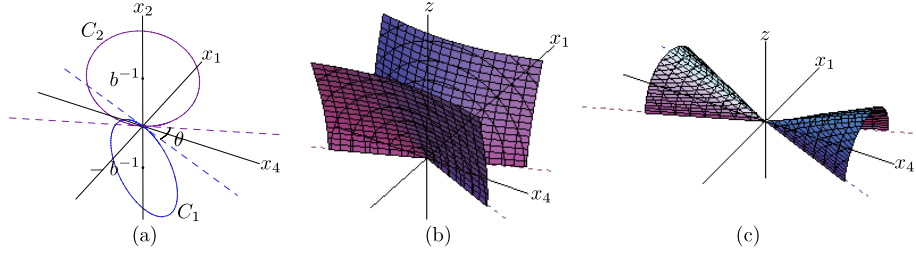}}
  \caption{(a)  The two circles forming the UV boundary in the inverted coordinates.  
 \\ (b) The two cusps with angle $\pi - \theta$ around the origin. 
 \\ (c) The two cusps with angle $\theta$ around the origin.}
\label{figquarkcirc}
\end{figure}
The two circles $C_1$ and $C_2$ are centered at $\mp b^{-1}$ in the
$x_2$-direction, respectively, and have radius $b^{-1}$ 
(see Fig.~\ref{figquarkcirc} a), so that they touch at the origin. More
precisely, the regions of the two straight lines corresponding to 
$-T \leq \tau \leq T$ are mapped into the regions of the circles
corresponding to $\xi$ in the range 
\begin{equation}
  \label{deficitangle}
-\pi + \xi_c \leq \xi \leq \pi - \xi_c \,, \quad 
0 \leq \xi_c \equiv \arcsin {bT \over T^2 + b^2/4} \leq {\pi \over 2} \,,
\end{equation}
with $\xi_c$ approximately equal to $\xi_c \sim b/T$ for large $T$. 
The regions $\tau \leq -T$ and $T \leq \tau$ of the straight lines
$L_1$ and $L_2$ are mapped into two arcs of the circles $C_1$ and
$C_2$, of opening angle $2\xi_c$. These arcs have a contact 
point at the origin, which corresponds to the points at infinity 
$\tau = \pm \infty$ of the lines $L_1$ and $L_2$. 
Around the contact point, where the arcs can be approximated by their tangents, 
one sees clearly the appearance of two crossing straight lines, 
which imply therefore the presence of a cusp-like region in the minimal surface 
(see Fig.~\ref{figquarkcirc} b,c).  

Indeed, two crossing lines give rise to two pairs of equal angles, 
namely $\theta$ and $\pi-\theta$. 
Since the boundaries correspond to fundamental Wilson lines, 
they have a definite orientation, and so
only one pair of angles can contribute. 
For quark-quark scattering ``at angle $\theta$'', 
the relevant minimal surface is defined in the original coordinates 
by a boundary formed by the two lines \eqref{lines}, and
so it is the pair of angles $\pi-\theta$ which gives a cusp
contribution to the corresponding minimal surface in the inverted
coordinates (see Fig.~\ref{figquarkcirc} b). 
In order to obtain the minimal surface 
for quark-antiquark scattering ``at angle $\theta$'', 
we have to reverse the orientation of one of the boundaries, as in
\eqref{antiq}, 
and so in this case it is the pair of angles $\theta$ 
which gives a cusp contribution (see Fig.~\ref{figquarkcirc} c). 
Of course, this corresponds to quark-quark scattering 
``at angle $\pi-\theta$'', as repeatedly pointed out.

The appearance of these cusps allows to improve the general expression
\eqref{resultJP} for the regularized area. For this sake, it is convenient
to work with the Legendre transform prescription of Ref.~\cite{DGO}, in
order to get rid of linear UV divergences. 
It is also convenient to work with the minimal surface
obtained in the new, inverted coordinates, which as we have explained
above gives the same result for the area up to an irrelevant constant. 
Let us split the IR-regularized, UV-subtracted area functional
evaluated on the minimal surface in the inverted coordinates, 
which we denote with $A^{\rm quark}_{\rm min}$
by introducing an intermediate time scale $\rho$:
\begin{align}
A^{\rm quark}_{\rm min}(\theta,b,T) &= 
A^{\rm quark}_{\rm fin}(\theta,b,\rho) + A^{\rm quark}_{\rm div}(\theta,b,T,\rho)\,, 
\label{JPareamin}\\
A^{\rm quark}_{\rm fin}(\theta,b,\rho) &= \int_{-\rho}^{\rho} d\tau \int_{-b/2}^{b/2} d\sigma \,{\cal L}  \,,
\label{JPareacutbis{a}}\\
A^{\rm quark}_{\rm div}(\theta,b,T,\rho) &= 
\biggl(\int_{-T}^{-\rho} + \int _\rho^T \biggr) d\tau
\int_{-b/2}^{b/2} d\sigma\,{\cal L} \,, \label{JPareacutbis{b}}
\end{align}
where for the sake of simplicity we did not write explicitly the
Legendre transform prescription terms.  
It is well-known that when the cut-off $T\to \infty$, the cusps of the
new geometrical boundary defined in \eqref{circles} (see also
Fig.~\ref{figquarkcirc})
provide a logarithmic divergence in the area functional~\eqref{JPareamin}.
\begin{figure}[t]
    \centering
    \epsfbox{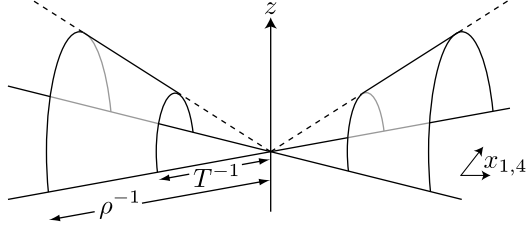}
    \caption{The contribution 
      to $A^{\rm quark}_{\rm div}(\theta,b,T,\rho)$ of the two cusps
      with angle $\theta$ at the origin.} 
\label{figcuspregion}
\end{figure}
By introducing an intermediate scale $\rho$, which is kept fixed in
the limit $T\to \infty$,  we are able to separate the divergent
contribution \eqref{JPareacutbis{b}}, which will be dominated by the cusp,
from a regular, finite part \eqref{JPareacutbis{a}}, see {\it e.g.}
Fig.~\ref{figcuspregion}.  
The scale $\rho$ is chosen to be large with respect to $b$ 
(and thus, after inversion,
$\rho^{-1}$ is small compared to the circle diameter in  \eqref{circles}), 
but it is otherwise arbitrary. 
Using conformal invariance, and exploiting 
the known properties of Wilson loop expectation values \cite{Mai,DGO,RY}, 
we have that 
\begin{equation}
  \label{qdiv}
A^{\rm quark}_{\rm div}(\theta,b,T,\rho) = 
2\Gamma_{\rm cusp}^E(\pi-\theta)\log{\rho \over T} + H(\theta,b/T,\rho/T)  \,,
\end{equation}
where $\Gamma^E_{\rm cusp}(\Omega)$ is a known function 
for Euclidean angle $0<\Omega<\pi$ calculated in Ref.~\cite{DGO}, 
and where $H(\theta,b/T,\rho/T)$ is finite in the limit $T\to\infty$. 
The factor of 2 is due to the fact that there are two cusp contributions. 
On the other hand, the term $A^{\rm quark}_{\rm fin}$ must take
the form $A^{\rm quark}_{\rm fin}(\theta,b,\rho) =
G(\theta,\rho/b)$. 
All in all, we have therefore 
\begin{equation}
  \label{qtot}
A^{\rm quark}_{\rm min}(\theta,b,T) = 2\Gamma_{\rm cusp}^E(\pi
-\theta)\log{\rho \over T} + H(\theta,0,0) + G(\theta,\rho/b) + o(T^0) \,,
\end{equation}
where $o(T^0)$ stands for terms which vanish in the limit
$T\to\infty$. As we have already said, the scale $\rho$ is a
fixed intermediate scale, allowing to singularize the cusp contribution to the area.
Now, since $\rho$ is arbitrary, it should disappear from
the right-hand side of Eq.~\eqref{qtot}, and this is possible only if 
\begin{equation}
  \label{qfinarb}
G(\theta,\rho/b) = -2\Gamma_{\rm cusp}^E(\pi -\theta)\log{\rho \over b} + {\hat G}(\theta) \,.
\end{equation}
This can be looked at also in a different way. 
We can take $\rho$ to be not an arbitrary ``external'' scale, 
but the one determined by the exact solution of the minimal surface problem,
that separates the region where the surface is well approximated by a
cusp solution from the rest. For dimensional reasons, it must be of
the form $\rho=b\,g(\theta)$, so that Eq.~\eqref{qfinarb} again follows.

In conclusion, comparing the minimal area \eqref{resultJP} with \eqref{qtot}, 
we can write\footnote{Terms of order $o(T^0)$ are
actually present in the full expression for $A^{\rm quark}_{\rm min}$ 
at finite $T$. 
This can be understood from the fact that in the limit $\theta\to 0$ 
we should recover the result for two parallel lines,
which is proportional to $T/b$. This would be the case if, for example, 
the exact expression were of the form 
$A^{\rm quark}_{\rm min}\sim \log\big[\exp\big({b/T \over (b/T)^2 +
\theta}\big) -1\big]$ at large $T$: while for $\theta\neq 0$
one would obtain $\sim\log (b/ T\theta)$, at $\theta=0$ one would
recover the linear divergence $\sim T/b$.}
\begin{equation}
  \label{qqcusplog}
\Phi_E(b/T,\theta) = 2\Gamma_{\rm cusp}^E(\pi-\theta)\log{b \over T} +
o(T^0)\,.   
\end{equation}
We notice that Eq.~\eqref{qqcusplog}\ contains only the contribution 
of the region around the contact point of the two circles, 
which is related by inversion to the region at infinity 
of the two straight lines. 
In other words, the $b,T$-dependent term Eq.~\eqref{qqcusplog} is determined
only by the initial and final data of quarks, and this reflects well
the link between the eikonal approximation and the dominance of the cusps. 
The relation between the cusp anomalous dimension and the
high-energy behavior of scattering amplitudes is a well-known fact,
but it is not evident {\it a priori} how this relation would
show up in the eikonal approach, where no cusp is present in the
initial setting, in the strong-coupling regime.\footnote{In the weak
coupling regime, the relation between the Wilson-line correlator and
the cusp anomaly has been investigated in perturbation theory in
Ref.~\cite{Kor}.} The result Eq.~\eqref{qqcusplog} thus provides a first 
nontrivial check for the viability of the eikonal approach.

On the other hand, the function $\Psi_E(\theta)$ in
Eq.~\eqref{resultJP} remains  
to be determined, which would require the exact solution of
the minimal surface problem, which is not available at the
moment. However, it is possible to go further and determine an 
interesting approximation by using the ``generalized helicoid''
ans{\" a}tz \eqref{helico}. It amounts to find a refined estimate  of the
intermediate scale $\rho$, in the ``natural'' sense discussed after
Eq.~\eqref{qfinarb}, isolating more precisely the (truncated) cusp
contribution.  

\subsection{The ``generalized helicoid'' ans{\" a}tz}

Let us go back to the regularized area functional \eqref{JParea} derived 
from the area functional \eqref{areafunc} with ``generalized helicoid''
ans\"atz \eqref{helico},  
discussed in Section 2. 
Following Ref.~\cite{JPi}, we make the change of variables
\begin{equation}
  \label{changevar}
\sigma' \equiv \sigma\sqrt{1+\biggl({\theta\tau \over b}\biggr)^2}\,,\quad
z'(\tau,\sigma')\equiv z(\tau,\sigma(\tau,\sigma'))\,, 
\end{equation}
which leads to the following expression for the area functional,
\begin{equation}
  \label{JPareachange}
  A^{\rm quark}_{\pi-\theta,b} = {\sqrt{\lambda} \over 2\pi}
  \int_{-T}^{T}d\tau
  \int_{-{b \over 2}\sqrt{1+\left({\theta\tau \over b}\right)^2}}^{{b
      \over 2}\sqrt{1+\left({\theta\tau \over b}\right)^2}}d\sigma   
	{1 \over z^2}\sqrt{1+ (\partial_\sigma z)^2+ 
	\Biggl( \partial_\tau z +{ \bigl({\theta\tau \over b}\bigr)
	\bigl({\theta\sigma \over b}\bigr) \over
        1+\bigl({\theta\tau\over b}\bigr)^2}\,\partial_\sigma
      z\Biggr)^2} \,,  
\end{equation}
where we have dropped the primes for simplicity. As we have already
remarked, the ans{\" a}tz \eqref{helico} is appropriate for quark-antiquark
scattering, as indicated by the subscript $\pi-\theta$.

It can be realized that, written in the form \eqref{JPareachange},
the ``generalized helicoid'' ans{\" a}tz admits interesting 
approximate while explicit solutions for both the large and  small
$|\tau|$ regions. 
\begin{itemize}
\item[i)] {\it Small-$|\tau|$ region, {\it i.e.}, $\theta|\tau|/b \ll
    1$} \\  
In this region  the corresponding contribution to the area
functional simplifies to 
\begin{equation}
  \label{JPareasmalltau}
A^{\rm quark}_{\pi-\theta,b}\big|_{{\rm small}\,\tau} 
= {\sqrt{\lambda} \over 2\pi} 
	\int_{-\delta{b /\theta}}^{\delta{b/\theta}} d\tau 
	\int_{-b/2}^{b/2} d\sigma 
	{1 \over z^2} \sqrt{ 1 +(\partial_\sigma z)^2 +(\partial_\tau z)^2} \,, 
\end{equation}
where $\delta$ is some small positive
number.\footnote{Eq.~\eqref{JPareasmalltau} can be obtained  
directly from Eq.~\eqref{JParea} in the small-$\tau$ region.
} 
This functional corresponds to
the area functional of a minimal surface with planar boundaries, where
the symmetries of the problem allow to write the solution in the form
\begin{equation}
  \label{boundplanar}
t = \tau\,,\quad 
x={\rm constant}\,,\quad 
y=\sigma\,,\quad 
z=z(\tau,\sigma)\,.  
\end{equation}
Moreover, in our case the boundary is made up of two segments of
parallel straight lines of length $2\delta{b /\theta}$ 
at a distance $b$, for which the solution is known~\cite{Mai}. 
The corresponding (regularized and UV-subtracted) area is
\begin{equation}
  \label{areasmalltau}
 A^{\rm quark}_{\rm min}(\pi-\theta,b,T)\bigr|_{{\rm small}\,\tau} 
	\simeq -{\sqrt{\lambda} \over 2\pi} c {2 \delta b \over
          \theta} {1\over b}  
  =  -{\sqrt{\lambda} \over 2\pi}{2c \delta \over \theta}\,,
\end{equation}
where the constant $c=8\pi^3/\Gamma^4(1/4)$ is the
coefficient in front of the (screened) coulombic potential~\cite{Mai}. 
One immediately sees that after analytic continuation 
this contribution is vanishing with energy, 
both for the quark-quark ($\theta\to \pi+i\chi$) and 
for the quark-antiquark ($\theta\to -i\chi$) $s$-channel scattering processes,
in the limit $\chi\to\infty$.
\item[ii)] {\it Large-$|\tau|$ region, {\it i.e.}, $\theta |\tau|
    /b\gg 1$} \\ 
Eq.~\eqref{JPareachange} is also suitable for an analytic solution 
in the large $\tau$ region. Neglecting 1 against $\theta\tau/b$, 
the area functional simplifies to
\begin{equation}
  \label{JParealargetau}
  A^{\rm quark}_{\pi-\theta,b}\bigr|_{{\rm large}\,\tau} =
  {\sqrt{\lambda} \over 2\pi}\biggl(\int_{-T}^{-\Lambda{b \over
      \theta}}+\int_{\Lambda{b \over\theta}}^{T}\biggr)d\tau 
  \int_{-{\theta |\tau| \over 2}}^{{\theta |\tau| \over 2}}d\sigma 
	{1 \over z^2}\sqrt{1+ (\partial_\sigma z)^2 
	+\Bigl( \partial_\tau z + {\sigma \over \tau}\partial_\sigma z\Bigr)^2}\,,
\end{equation}
where $\Lambda$ is some large number. 
Away from the boundary, where $|\sigma / \tau|$ is small, 
Eq.~\eqref{JParealargetau} can be further approximated as
\begin{equation}
  \label{JParealargetaubis}
A^{\rm quark}_{\pi-\theta,b}\bigr|_{{\rm large}\,\tau} = {\sqrt{\lambda} \over 2\pi}
  \biggl(\int_{-T}^{-\Lambda{b \over\theta}}
  +\int_{\Lambda{b \over\theta}}^{T}\biggr)d\tau
  \int_{-{\theta |\tau| \over 2}}^{{\theta |\tau| \over 2}}d\sigma\,  
{1 \over z^2}\sqrt{1+ (\partial_\sigma z)^2+ \left(
 \partial_\tau z\right)^2}\,.  
\end{equation}
We have again to deal with a minimal surface with planar boundary,
which this time consists of two segments of straight lines at an angle $\theta$,
\begin{equation}
  \label{straightbound}
  \sigma_\pm(\tau)=\pm {\theta\tau \over 2} \,,
\end{equation}
with $|\tau|\in [\Lambda b /\theta, T]$. The solution is
immediately seen to be made up of two parts, each corresponding to a
piece of the solution for a cusp of angle $\theta$ ({\it cf.}
Fig.~\ref{figcuspregion}),  
and the resulting (regularized and UV-subtracted) area is 
\begin{equation}
  \label{arealargetau}
  A^{\rm quark}_{\rm min}(\pi-\theta,b,T)\bigr|_{{\rm large}\,\tau} = 2\Gamma_{\rm
  cusp}^E(\theta)\log{\Lambda b \over T \theta}\,.
\end{equation}
This result is in agreement with
the general form \eqref{resultJP} for the minimal area,\footnote{We note in
passing that this agreement is for two reasons 
in favor of our choice of using the
ans{\" a}tz \eqref{helico}: we obtain the cusp contribution predicted by our
general considerations, and also the $\theta$ factor inside the
logarithm which is expected, after analytic continuation, by
comparison with the Alday-Maldacena and the Barnes-Vaman amplitudes.} 
and moreover allows    
to determine the ``natural'' choice of a $\theta$-dependent scale 
$\rho \sim \Lambda b /\theta$, discussed after Eq.~\eqref{qfinarb}, 
which separates the near-cusp region from the rest 
in the inverted coordinates.\footnote{Note that the divergence in 
$A^{\rm quark}_{\rm min}(\pi-\theta,b,T)\bigr|_{{\rm large}\,\tau}$
comes from the 
large-$\tau$ region, {\it i.e.}, far away from the cusp appearing in the
original coordinates, which corresponds to the near-cusp region in the
inverted coordinates.
} 
Indeed, up to the constant $\Lambda$, whose precise value 
cannot be determined at the present stage, 
we have that $\rho \propto b/\theta$. 
The factor $1/\theta$ could not be predicted
with the general arguments of the previous subsection: 
its important role will become clear after analytic continuation 
to Minkowski space.
Let us finally remark that Eq.~\eqref{arealargetau} gives  also an
estimate of the 
function $\Psi_E(\theta)$ in Eq.~\eqref{resultJP}:
\begin{equation}
  \label{psie}
\Psi_E(\theta) \sim  2\Gamma_{\rm
  cusp}^E(\theta)\log{\Lambda \over \theta}\,,  
\end{equation}
up to the term Eq.~\eqref{areasmalltau}, which as we have explained gives a
vanishing contribution after analytic continuation, and up to possible
contributions from the intermediate region
$\tau\theta/b\in[\delta,\Lambda]$, 
as well as from the region $\sigma \simeq (\theta /2) \tau$. 
In a sense,\footnote{The above-mentioned contributions are
not expected to change too much the results above, Eqs.~\eqref{arealargetau} and
\eqref{psie}: the intermediate-$\tau$ region should somehow interpolate
between Eqs.~\eqref{areasmalltau} and \eqref{arealargetau}, while the
near-boundary region 
basically contributes the UV-divergent $1/\epsilon$ term which
is removed by the Legendre transform prescription, and so the exact
behavior of the surface in this region should not affect too much the
result. Although these issues require further work to be clarified, we
believe that these terms lead to contributions subleading in energy
(or at most of order ${\cal O}({\chi})$) after analytic
continuation, which can therefore be safely neglected without altering
our conclusions.
} 
the constant $\Lambda$ stands for our ignorance 
about the $b,T$-independent term $\Psi_E(\theta)$. 
\end{itemize}

We are now ready to perform the analytic continuation. 
Neglecting subleading contributions, and considering for definiteness
the quark-quark $s$-channel, so that the relevant analytic
continuation reads\footnote{See Eq.~\eqref{ancontrel}. Note that we
  are working with $A^{\rm quark}_{\rm min}(\pi-\theta,b,T)$.}  
\begin{equation}
  \label{ancontuchan}
\theta \to \pi + i \chi\,, \quad T\to iT\,,  
\end{equation}
with $\chi \sim \log(s/M^2)$, $s>0$, we obtain
\begin{equation}
  \label{areamink}
A^{{\rm quark},\,s}_{{\rm min},\,M}(\chi,b,T) 
= 2 \Gamma_{\rm cusp}(\chi)\log{\Lambda b \over T \chi e^{i\pi}\bigl(1 +
  e^{-i{\pi\over 2}}(\pi/\chi) \bigr)}
= 2 \Gamma_{\rm
cusp}(\chi)\log{ b \over T \chi} + {\hat\Psi}^{s}_{M}(\chi)\,,
\end{equation}
where we have used $\Gamma_{\rm cusp}^E(\pi + i\chi)= \Gamma_{\rm
cusp}(\chi)$~\cite{Kr}.   
Taking the limit $\chi\to \infty$, we obtain for the   
$b,T$-dependent term and for the leading $\chi$-dependence
\begin{equation}
  \label{areaminkbis}
A^{{\rm quark},\,s}_{{\rm min},\, M}(\chi,b,T) = 
-{f(\lambda) \over 2}\chi  \log{ b \over T \chi} + {\cal O}(\chi)\,,
\end{equation}
where we have used Eq.~\eqref{cuspmink}, 
which also implies that the auxiliary function
${\hat\Psi}^{s}_{M}(\chi)={\cal O}(\chi)$ in \eqref{areaminkbis}. 

The $u$-channel quark-quark amplitude, 
\begin{equation}
  \label{quchamp}
{\tilde \CA}^{{\rm quark},\,u}(\chi,b,T) \equiv 
\exp\bigl[-A^{{\rm quark},\,u}_{{\rm min},\,M}(\chi,b,T)\bigr] 
\quad \bigl(= \tilde\CA^{{\rm quark}}_E(\pi+i\chi,b,iT)\bigr) \,,  
\end{equation}
that we shall use in the next subsection for the comparison 
with the results of the Alday-Maldacena approach, 
is obtained by means of the crossing-symmetry relations \eqref{crossrel},
$i.e.$, through the analytic continuation\footrep 
\begin{equation}
  \label{ancontuchanu}
\theta \to - i \chi\,, \quad T\to iT\,,  
\end{equation}
with $\chi \sim \log(-s/M^2)$, $u\sim -s>0$, which yields
\begin{equation}
  \label{areaminku}
A^{{\rm quark},\,u}_{{\rm min},\, M}(\chi,b,T) = 2 \Gamma_{\rm
cusp}(i\pi-\chi)\log{\Lambda b \over T \chi}
= 2 \Gamma_{\rm cusp}(i\pi-\chi)\log{ b \over T \chi} +
{\hat\Psi}^{u}_{M}(\chi) \,.
\end{equation}
Although the exact value of $\Gamma_{\rm cusp}(i\pi-\chi)$ is not yet known,
we expect that its large-$\chi$ behavior coincides with that of
$\Gamma_{\rm cusp}(\chi)$ (this is actually the case in perturbation
theory~\cite{Po,Ko}),  
so that in the limit $\chi\to \infty$ the leading term
reads\footnote{The same high-energy limit is obtained by means of the usual
 analytic continuation of the area \eqref{areaminkbis} in terms of the
 Mandelstam variables,  $s\to e^{-i\pi}u$. 
}
\begin{equation}
  \label{areaminkbisu}
A^{{\rm quark},\,u}_{{\rm min},\, M}(\chi,b,T) = 
-{f(\lambda) \over 2}\chi  \log{ b \over T \chi}+ {\cal O}(\chi) \,,  
\end{equation}
which also implies that the auxiliary function in the
$u$-channel verifies ${\hat\Psi}^u_{M}={\cal O}(\chi)$. 

Our result \eqref{areaminkbisu} calls for a comment 
related to the initial approach of  Ref.~\cite{JPi}. 
In Ref.~\cite{JPi}, the functionals \eqref{JParea} and
\eqref{JPareachange} were the 
starting point for an approximate evaluation of the area of the
minimal surface. 
In particular, the aim of the authors was to determine the $T$-independent, 
IR-finite contribution to the area. 
To this extent, neglecting the non-diagonal terms 
in $\partial_\tau z$, $\partial_\sigma z$ in Eq.~\eqref{JPareachange}, 
they performed the angular part only of the analytic continuation, 
{\it i.e.}, $\theta\to -i\chi$ (see Eqs.~(31){--}(34) in Ref.~\cite{JPi}). 
The $T$-independent part of the resulting functional turned out to be
the area $A_{\rm ellipse}$ of a simpler minimal surface, 
living in Euclidean $AdS_5$, and having as boundary a half-ellipse 
of width $b/\chi$ and height $b$. 
Finally, the approximate evaluation of $A_{\rm ellipse}$ 
led to the following result:
\begin{equation}
  \label{ellipse}
A_{\rm ellipse} = -2\Gamma_{\rm cusp}^E\Bigl({\pi \over 2}\Bigr)\log{M_B b \over \chi} 
- {\sqrt{\lambda} \over 2\pi}{ c\pi \over 4} \chi\,,
\end{equation}
where $\Gamma_{\rm cusp}^E(\pi/2)$ is the Euclidean cusp anomaly 
calculated in Refs.~\cite{DGO,Kr} 
and $c$ is the same constant as that in Eq.~\eqref{areasmalltau}.
The scale $M_B^{-1}$ is the inverse mass of the $W$-bosons playing the
role of ``Euclidean quarks'', see Section 2, and corresponds to the
position of the D3-brane which acts as UV cutoff.

Our present study gives a different and improved answer 
to the problem initiated by Ref.~\cite{JPi}, 
as shown by comparing \eqref{areaminkbisu} and \eqref{ellipse}. 
In this paper we have gone beyond the approximations made in Ref.~\cite{JPi},
whose results suffer from the limited knowledge on minimal surface solutions 
for scattering amplitudes available at that time, 
in particular regarding the geometry relevant for quark-quark
scattering in the eikonal approach. 
The key point here are the non-diagonal terms 
in the area functional~\eqref{JPareachange}, 
which cannot be neglected in the region considered in Ref.~\cite{JPi}. 
Though functionally similar to \eqref{areaminkbisu} (by the interchange
of $\Gamma_{\rm cusp}^E(\pi/2)$ with $\Gamma_{\rm cusp}(i\pi-\chi)$), 
the expression \eqref{ellipse} does not contribute a non-factorizable factor 
to the amplitude. 
Moreover, the expression \eqref{ellipse} shows the appearance in the 
logarithmic term of the UV-cutoff $M_B$. 
As discussed above, $M_B$ should drop from the area 
when UV divergencies have been removed.

\subsection{Eikonal {\it vs.} Alday-Maldacena approach}

Let us finally compare our results for quark-quark scattering,
obtained in the eikonal approach, with the ones obtained for
gluon-gluon scattering using the Alday-Maldacena solution. Since we
are interested in the high-energy Regge behavior of the amplitude,
this is a sensible comparison to be made, due to the universality
property discussed in Section~4.

For convenience, we rewrite here the $u$-channel
quark-quark scattering amplitude obtained with the eikonal approach
(see Eqs.~\eqref{quchamp} and \eqref{areaminkbisu}),  
\begin{equation}
\label{areaminkteru}
-\log {\tilde \CA}^{{\rm quark},\,u}_{\rm eikonal}(\chi,b,T) =
-{f(\lambda) \over 2}\chi  \log{ b \over T \chi} + {\cal O}(\chi) 
= -{f(\lambda) \over 2}{\hat \chi}  \log{ b \over T {\hat \chi}}
+\CO({\hat \chi}) \,, 
\end{equation}
where we used $\chi = {\hat \chi} +\log(m^2/M^2)$, see Eqs.~\eqref{Quarkchi}
and \eqref{Gluonchi},  
and also the gluon-gluon scattering amplitude in
impact-parameter space and in the Regge limit obtained with the
Alday-Maldacena approach, Eq.~\eqref{Logampl},
\begin{equation}
  \label{Logamplbis}
-\log {\tilde \CA_{\rm AM}^{{\rm gluon}}}({\hat \chi},b,m) = -{f(\lambda) \over
2}{\hat \chi} \log {mb \over {\hat \chi}}  
	+{\hat \chi}\biggl[{f(\lambda) \over 2}\biggl(\log {f(\lambda)
        \over 2e}-i{\pi \over 2}\biggr) -{g(\lambda) \over 4}\biggr] +
        \cdots \,,
\end{equation}
where we have made explicit the dependence of the amplitude on the IR
regulator $m$, and we have specified which approach has been used with
appropriate subscripts.\footnote{The $u$-channel quark-antiquark
scattering amplitude, 
$i.e.$, the $s$-channel quark-quark amplitude
$\tilde \CA^{{\rm quark},\,s}_{\rm eikonal}(\chi,b,T)$ 
corresponding to
Eq.~\eqref{areaminkbis}, is exactly of the same form of
Eq.~\eqref{areaminkteru}, so 
our conclusions apply to this case as well.}

Examining  the expression for the quark amplitude \eqref{areaminkteru}
following the order in the expansion of the exact expression
\eqref{Logamplbis} for the gluon one, the following consequences can be
drawn: 
\begin{itemize}
\item[i)] {\it First term} \\ 
The first term exactly
coincides with the leading term \eqref{amplcusp} obtained in
the case of gluon-gluon scattering from the Alday-Maldacena
solution, up to a rescaling $T\to m^{-1}$, {\it i.e.}~up to a shift 
$$-{f(\lambda) \over 2}{\hat \chi}  \log {mT}={\cal O}({\hat \chi})$$ 
which plays a role at next to leading order only. 
Looking back to the discussion of the exact gluon-gluon amplitude
\eqref{amptotal}, we noticed that the first term in its impact-parameter
representation Eq.~\eqref{Logampl}, coinciding with \eqref{amplcusp} at high
energy, was at the origin of the Regge nature of the amplitude, and   
of the  $t$-dependent part of the Regge trajectory
\eqref{trajectory}. This implies  
that the quark-quark (and also quark-antiquark) scattering amplitude
is of Regge type, and that the $t$-dependent part of the Regge
trajectory is indeed the same obtained in the Alday-Maldacena
approach. \\
Hence the main conclusion is that the same  Regge factor
$(-s)^{-(f(\lambda)/4)\log (-t)}$ appears in the
$(s,t)$-representation of both amplitudes. This corresponds to the
fact that both amplitudes in impact-parameter space contain
the same term, {\it i.e.}, $2\Gamma_{\rm cusp}({\hat \chi}) \log[{\rm
  (mass)} \cdot b/{\hat \chi}]$.   
In particular, we notice that the $t$-dependent part of the Regge
trajectory comes entirely from the non-factorizable term $2\Gamma_{\rm
cusp}({\hat \chi}) \log[{\rm (mass)} \cdot b]$, which has been
obtained through the general considerations of Section 5.1 (see
Eq.~\eqref{qqcusplog}). This is therefore a robust result, independent of the
approximations performed in Section 5.2. 
It is also interesting to note that the leading term of order
${\hat \chi}\log{\hat \chi}$ in the factorized 
${\hat \chi}$-dependent part appears to be the same, while coming from
seemingly different origin in the two cases: in the quark amplitude it
comes from a refined evaluation of the cusp contribution, see
{\it e.g.}~\eqref{psie}, with the ``generalized helicoid'' ans{\" a}tz,
while in the 
gluon case it comes from the Fourier transform factor \eqref{Gchi} after
analytic continuation. As we have already remarked, this term is
essential in order to obtain an amplitude of Regge type.
\item[ii)] {\it Second term} \\
The  ${\cal O}({\hat \chi})$ term in  \eqref{areaminkteru} 
is compatible with Regge behavior. 
At the 
present stage we are not able to find a precise evaluation of this
term, which could be obtained from the full solution of the minimal
surface problem. However, as it has already been shown for the gluon
case (see Eqs.~\eqref{amptotRegge}{--}\eqref{residue}), 
it may affect only the factorized part of the amplitude, 
which depends on the regularization scheme. 
In particular, the $t$-dependent factorized term of the amplitude is
not expected to be universal, but to depend on the species of
the scattering particles.  
\item[iii)] {$f\log f$ {\it term}} \\
The $f\log f$ term in \eqref{Logamplbis} may seem puzzling at first,
since no term of this kind can be found in the expression for the area
of the minimal surface in the eikonal approach. However, its origin
becomes evident when one recalls that the radial coordinates $r$ and
$z$ used in the two approaches are related as
$r=R^2/z=\sqrt{\lambda}/z$, so that an appropriate conversion factor
has to be used when comparing the IR cutoffs. This is particularly
clear if one uses the radial cutoff $r_c$, which, as we have discussed
above in Section 4, is related to the cutoff $m$ in the dimensional
regularization scheme as $m = \tilde m\sqrt{2/e} = {r_c /(2\pi
\sqrt{e})}$. In turn, $r_c$ can be expressed as $r_c=R^2/z_c
= \sqrt{\lambda}/z_c$ in terms of the radial coordinate $z$, with
$z_c\to\infty$ when the IR regularization is removed, which is
appropriate for comparing the Alday-Maldacena result with the eikonal
approach. Expressing the leading term in Eq.~\eqref{Logamplbis} in
terms of this new cutoff, we see that the $f\log f$ term actually gets
cancelled, and we obtain the expression 
\begin{equation}
  \label{amplcuspzco}
-\log {\tilde \CA}_{\rm AM}^{\rm gluon}({\hat \chi},b,z_c) 
 = -{f(\lambda)\over 2}\hat\chi \log {b \over z_c{\hat \chi}} 
-{f(\lambda)\over 2}\hat\chi\bigg(1-\log\sqrt{2} + i{\pi \over
2}\bigg) + \cdots
\,.   
\end{equation}
The cutoff $z_c$ can now be identified with $T$, up to numerical
factors which affect only the regularization-scheme dependent part of
the amplitude. In other words, the shift proportional to $\log {mT}$,
discussed above in point i), naturally contains the appropriate
``counterterm'' which makes $f\log f$ drop from the complete
expression. 
\item[iv)] {\it Gluon-gluon scattering} \\
To conclude this section, we want to briefly discuss how the technique
applied above to quark-quark scattering is extended to the case of
gluon-gluon scattering. Recall from Section 2 the expressions
\eqref{ggocteta} and \eqref{ggoctets} for the ``octet'' component of
the amplitude. Using large-$N_c$ factorization and minimal surfaces,
we have to leading order in $N_c$
\begin{align}
\langle {\rm tr} W_1^{\dag} {\rm tr} W_2^{\dag} {\rm tr}[W_1 W_2]
\rangle &\sim \langle {\rm tr} W_1^{\dag}\rangle\langle {\rm tr}
W_2^{\dag} \rangle\langle {\rm tr}[W_1 W_2] \rangle \nonumber \\ &
\phantom{aaaaaaaaaaaaaaai} 
\sim N_c^3\biggl(1+e^{-A^{\rm quark}_{\rm
min}(\theta,b,T)}\biggr) \label{gluonloops} \\
\langle {\rm tr} W_1 {\rm tr} W_2^{\dag} {\rm tr}[W_1^{\dag} W_2]
 \rangle &\sim \langle {\rm tr} W_1\rangle\langle  {\rm tr}
 W_2^{\dag}\rangle \langle  {\rm 
 tr}[W_1^{\dag} W_2] \rangle \nonumber \\ & \phantom{aaaaaaaaaaaaaaai} 
\sim N_c^3\biggl(1+e^{-A^{\rm quark}_{\rm
min}(\pi-\theta,b,T)}\biggr) \label{gluonloopsbis} \\
\langle |{\rm tr} W_1 {\rm tr} W_2|^2 \rangle &\sim |\langle {\rm tr}
 W_1 \rangle \langle {\rm tr}
 W_2 \rangle|^2  \sim N_c^4\,, \label{gluonloopster}
\end{align}
and moreover $Z_V \sim  {1\over N_c^2}\langle  \tr W_i \rangle
\langle  \tr W_i^{\dag} \rangle 
 \sim 1 $. We therefore conclude that
 \begin{align}
\CA^{gg}_{N_c^2-1,A}  \sim & N_c^3 \biggl(e^{-A^{\rm quark}_{\rm
min}(\theta,b,T)} - e^{-A^{\rm quark}_{\rm
min}(\pi-\theta,b,T)}\biggr) 
\nonumber \\
& \mathop\to_{\theta\to -i\chi, T\to iT} ~~~ N_c^3\biggl( {\tilde
\CA}^{{\rm quark},\,s}_{\rm eikonal}(\chi,b,T)- 
{\tilde \CA}^{{\rm quark},\,u}_{\rm eikonal}(\chi,b,T)\biggr)\,,
 \label{ggoctetamins} \\ 
\CA^{gg}_{N_c^2-1,S}  \sim & N_c^3 \biggl(e^{-A^{\rm quark}_{\rm
min}(\theta,b,T)} + e^{-A^{\rm quark}_{\rm
min}(\pi-\theta,b,T)}\biggr)
\nonumber \\
& \mathop\to_{\theta\to -i\chi, T\to iT} ~~~ N_c^3\biggl( {\tilde
\CA}^{{\rm quark},\,s}_{\rm eikonal}(\chi,b,T)+ 
{\tilde \CA}^{{\rm quark},\,u}_{\rm
  eikonal}(\chi,b,T)\biggr)\,. \label{ggoctetsmins} 
 \end{align}
As already anticipated in the Introduction, the calculation in the
gluon-gluon case reduces basically to that of the quark-quark
case. Moreover, it is evident from Eqs.~\eqref{ggoctetamins} and
\eqref{ggoctetsmins} that the high-energy behavior is the same in the two
cases. In particular, together with the expressions \eqref{areaminkbis} and
\eqref{areaminkbisu}{--}\eqref{areaminkteru} for the high energy
behavior of the quark 
amplitudes, this result shows that also the ``octet'' component of the
gluon-gluon scattering amplitude is of Regge type, with the same gluon
Regge trajectory as in the quark-quark case, and therefore with the same
$t$-dependent part of the trajectory found by Alday and
Maldacena. As a final remark, we want to stress the fact that
universality is shown in a simpler way in the eikonal approach, thanks
to the fact that the basic object in the computation of the scattering
amplitude is the correlation function of the same Wilson lines,
differing only for the representation in which they are taken. 
\end{itemize}

\nsection{Summary, comments  and  outlook}

In this work we have investigated the Regge behavior of high-energy
amplitudes in $\CN \!=\! 4$ supersymmetric Yang-Mills theory at strong
coupling, using the AdS/CFT correspondence in two different ways. 
For this sake we have analyzed
these amplitudes in the dual gravity theory, where they are obtained
as the (regularized) area of minimal surfaces in  Minkowskian AdS and
hyperbolic (or ``Euclidean AdS'') backgrounds. We summarize here the
main points.
\begin{itemize}
\item[i)] 
In order to make easier the comparison with the eikonal approach, the
Alday-Maldacena four-gluon amplitude~\cite{AM}, obtained from a minimal
surface in Minkowskian AdS, has been put in a Regge
form~\cite{NS,DKS}, see formulas \eqref{amptotRegge} and
\eqref{trajectory}, namely 
\begin{equation}
  \label{amptotReggefinal}
\CA (s,t)  \equiv  \beta(t)\biggl({-s \over m^2}\biggr)^{\alpha(t)} 
\propto
C_\epsilon\biggl({-t \over m^2}\biggr)^{{g(\lambda) 
\over 4}-1}\biggl({-s \over m^2}\biggr)^{\alpha(t)}\ ,
\end{equation}
where $m$ is an IR cut-off and $\alpha(t)$ is the {\it Regge
  trajectory}, 
\begin{equation}
  \label{trajectoryfinal}
\alpha(t) = -{f(\lambda) \over 4}\log{-t \over m^2} +
  {g(\lambda)\over 4} +1 \,,    
\end{equation}
$C_\epsilon$ is a regularization-dependent constant, 
and the functions $f(\lambda)$ and $g(\lambda)$ have been defined 
in Eqs.~\eqref{cuspanomdim}. It is known that the trajectory is identified
with the gluon Regge trajectory~\cite{DKS}, corresponding to the exchange of
gluon quantum numbers between the colliding particles. The same
$t$-dependent part of the Regge trajectory is found in the quark-quark
elastic scattering calculation of Ref.~\cite{BV}, in accordance with the
expected universality of the Regge behavior. \\
In order to compare this to the results obtained in the eikonal
approach, we have also studied the corresponding impact-parameter
representation, where the Regge nature of the amplitude is encoded in
the leading factor $(mb/\chi)^{-{f(\lambda) \over 2}\chi}$. 
\item[ii)] We have computed the ``octet''-exchange component of the quark-quark
and quark-antiquark 
elastic amplitude at high-energy in the {\it impact-parameter}
representation, by using the eikonal method in hyperbolic space~\cite{JPi}.
This amounts to consider the (regularized) minimal surface
corresponding to a ``generalized helicoid'' in hyperbolic space,
{\it i.e.}~the surface bounded by two straight lines at the Euclidean boundary. 
By performing a conformal transformation, we have 
shown that the minimal area is dominated by the contribution of two
identical cusps, which leads to the same $t$-dependent part
$-(f(\lambda)/4)\log(-t/m^2)$ of the Regge trajectory
\eqref{trajectoryfinal}, where $f(\lambda) / 4$ is the coefficient
of the cusp anomalous dimension in Minkowski space \eqref{cuspmink}. This
shows the compatibility between the two {\it a priori} very different
approaches, making us confident in the viability of the eikonal method
in the physically interesting case of QCD, where the Alday-Maldacena
method is not available.   
\item[iii)] We have also computed the ``octet''-exchange component of the
gluon-gluon elastic scattering amplitude in the eikonal approach, which boils
down to a linear combination of the corresponding results for quark-quark
and quark-antiquark scattering. In this way we have shown universality
of the Regge behavior in the framework of the eikonal method, which is
obtained in a simpler way than in the Alday-Maldacena approach. 
\end{itemize}

Let us finally propose an outlook on open questions.

The Alday-Maldacena solution \eqref{solorig} in the position space 
is described in terms of complex coordinates, namely the target space
is extended to the complexified $AdS_5$.  
We have generated the new minimal surface \eqref{solEuc} from
Eqs.~\eqref{solorig}  
by performing the Wick rotation of the time coordinate of $AdS_5$, as
well as of the two world-sheet coordinates.  
The resulting surface is embedded into the ordinary Euclidean $AdS_5$,
and its boundary lies in the UV region ($i.e$, near the boundary $z=0$
of Euclidean $AdS_5$), while on the other hand the surface described by the
Alday-Maldacena solution has its boundary in the IR region of
(Minkowskian) $AdS_5$. We have found that the UV boundary of our
solution is a set of multiple helices; in particular, in the
forward Regge limit, $-s \to \infty$ with $-t$ fixed, the boundary
reduces to a double helix. 
This hints to the existence of a helicoid structure common to the two
approaches, which however results in different surfaces in the
Euclidean AdS background, in some sense ``dual'' under interchange of
two boundaries of a truncated helicoid, see
Fig.~\ref{figwilsonloop}. Further studies are required in order to
fully understand this similarity.  

In order to perform the comparison between the two approaches, we have
calculated both scattering amplitudes in the same Minkowskian impact
parameter $(b,s)$-representation. The amplitudes consist of
non-factorizable and factorizable parts with respect to $b$ and $s$. 
The area of both minimal surfaces contains 
the same leading non-factorizable term 
$2\Gamma_{\rm cusp}({\hat \chi})\log (m b/{\hat \chi})$ with $\hat\chi
\sim \log s$ (under the
rescaling $m \to T^{-1}$, and up to subleading terms), 
compare \eqref{areaminkteru} with \eqref{Logamplbis}, which leads 
to the same $t$-dependent part of the Regge trajectory, as we have
already remarked. The exact subleading term has not yet been obtained
in the eikonal approach, which requires the exact solution of the
``generalized helicoid'' problem. 

An important point concerns the physical relevance of the
subleading terms in $\chi$. Such terms are not known in the eikonal
approach, due to the lack of an exact solution for the minimal surfface.
However, such terms are finally involved in the regularization-scheme
dependence, and one may ask what is their physical relevance. Stated
differently, would we know more about the physics of scattering
amplitudes if we knew those terms exactly? This is an open problem for
future investigations. 

\subsection*{Acknowledgments}
MG and SS are grateful to IPhT, CEA-Saclay for hospitality. 
MG was supported by MICINN under the CPAN project CSD2007-00042 from the
Consolider-Ingenio2010 program, as well as under the grant
FPA2009-09638.

\end{document}